\documentclass[twocolumn]{aastex701}
\usepackage{CJK}
\usepackage{{amsmath}}

\begin{document}
\begin{CJK*}{UTF8}{gbsn}

\title{Unlocking the QPE Mystery: Star-Disk Collisions in Realistic AGN Disks}

\author[0000-0003-3616-6822]{Zhaohuan Zhu (朱照寰)}
\affiliation{Department of Physics and Astronomy, University of Nevada, Las Vegas, 4505 South Maryland Parkway, Las Vegas, NV 89154, USA}
\affiliation{Nevada Center for Astrophysics, University of Nevada, Las Vegas, 4505 South Maryland Parkway, Las Vegas, NV 89154, USA}
\email{zhaohuan.zhu@unlv.edu} 

\author[0000-0003-2868-489X]{Xiaoshan Huang (黄小珊)}
\affiliation{California Institute of Technology, TAPIR, Mail Code 350-17, Pasadena, CA 91125, USA}
\email{xshuang@caltech.edu} 

\author[0000-0002-2624-3399]{Yan-Fei Jiang (姜燕飞)}
\affiliation{Center for Computational Astrophysics, Flatiron Institute, New York, NY 10010, USA}
\email{yjiang@flatironinstitute.org} 

\author[0000-0002-7276-3694]{Shunquan Huang (黄顺权)}
\affiliation{Department of Physics and Astronomy, University of Nevada, Las Vegas, 4505 South Maryland Parkway, Las Vegas, NV 89154, USA}
\affiliation{Nevada Center for Astrophysics, University of Nevada, Las Vegas, 4505 South Maryland Parkway, Las Vegas, NV 89154, USA}
\email{huangs18@unlv.nevada.edu}

\begin{abstract}

Quasi-periodic eruptions (QPEs) are luminous, recurring soft X-ray outbursts observed in the nuclei of low-mass galaxies. They display two remarkable trends: outburst durations are $\sim$10--20\% of the recurrence timescale, and longer bursts are more luminous. A promising theory that naturally explains the quasi-periodicity invokes collisions between a star on an extreme mass-ratio inspiral (EMRI) orbit and the accretion disk around the supermassive black hole. However, it remains unclear how this model reproduces the observed trends. We therefore carry out two-dimensional, multi-frequency radiation hydrodynamic (RH) simulations of star--disk collisions. Crucially, we adopt a more realistic circumnuclear disk structure from previous Radiation MHD simulations of sub-Eddington accretion disks. We find that the thick, puffed-up disk atmosphere, extending to $z/r\sim1$, causes 
different portions of the bow shock to break out at different times, producing 
prolonged thermal emission as the shock emerges through the breakout 
surface at $z/r\sim0.7$. The X-ray flare duration is set by the shock propagation time through the optically thick disk---a $\sim$10\% of the orbital timescale, reproducing the observed duty cycle. A more oblique star--disk interaction yields a longer, more luminous flare. The realistic AGN disk models also exhibit a surface density $\Sigma\propto r^2$, giving a collisional energy $E\propto P^{2/3}$ that may explain the luminosity--period trend, especially for the weaker QPEs. Overall, we suggest that a more realistic circumnuclear disk structure can explain several observed QPE trends---and QPEs may, in turn, constrain the disk structure.

\end{abstract}



\section{Introduction} \label{sec:intro}
Quasi-periodic eruptions (QPEs) are recurring soft X-ray transients 
\citep{miniutti2019nine,giustini2020x,arcodia2021x,chakraborty2021possible,Quintin2023,
nicholl2024quasi,chakraborty2025discovery,HernandezGarcia2025} with recurrence times 
ranging from a few hours to several days and individual flare durations from sub-hour 
to a few days. The typical flare duration is $\sim$10--20\% of the recurrence time 
\citep{nicholl2024quasi,chakraborty2025discovery}. Assuming isotropic emission, flare 
luminosities reach $L_\mathrm{X} \sim 10^{41\text{--}43}\,\mathrm{erg\,s^{-1}}$, and 
their soft X-ray spectral energy distributions (SEDs) are well described by blackbody 
spectra with characteristic temperatures $kT \sim 100$--$200\,\mathrm{eV}$. During 
each flare, the emission hardens near peak and softens during the decline 
\citep{arcodia2024more,arcodia2025srg}. Between flares, QPE sources exhibit quiescent 
emission at $L_\mathrm{X} \sim 10^{41}\,\mathrm{erg\,s^{-1}}$ and $kT \sim 50\,\mathrm{eV}$. 
To date, no variability associated with QPE flares has been detected at other wavelengths 
\citep[e.g.][]{wevers2025time,goodwin2025radio}, except for ANSKY with a potential UV counterpart \citep{Guo2026}.

QPEs are preferentially found in the nuclei of low-mass galaxies \citep{Wevers2022}, 
pointing to an association with supermassive black holes (SMBHs) at galactic centers. 
Some QPE hosts show evidence of past tidal disruption events (TDEs) 
\citep{nicholl2020outflow,nicholl2024quasi,newsome2024mapping,chakraborty2025discovery}, 
and one case is associated with a previously active galactic nucleus 
\citep{HernandezGarcia2025,hernandez2025nicer,zhu2025ultraviolet,Guo2026}. 
Pre-existing circumnuclear gas may be present even in TDE-associated cases 
\citep{short2023delayed,newsome2024mapping,chakraborty2025discovery}, suggesting a 
connection between QPEs and low-luminosity accretion disks around SMBHs.

Several mechanisms have been proposed to explain the quasi-periodicity: disk 
instabilities \citep{miniutti2019nine,raj2021disk,pan2022disk,kaur2023magnetically}, 
episodic accretion driven by mass transfer from a stellar companion 
\citep{zalamea2010white,king2020gsn,king2022quasi,zhao2022quasi,metzger2022interacting,
krolik2022quasiperiodic,linial2023unstable,lu2023quasi}, and repeated perturbations 
from an orbiting companion object 
\citep{xian2021x,linial2023emri+,franchini2023quasi,tagawa2023flares,yao2025star,
vurm2025radiation,dodd2025perturbing,huang2025multi,suzuguchi2025quasi,jiang2025embers,2026HuangS,liu2026quasi}.

Among these, the companion--disk interaction model naturally reproduces the recurrent 
flares \citep{linial2023emri+,tagawa2023flares}. QPE timing places stringent constraints on the 
perturber's orbital radius: since the companion crosses the disk twice per orbit, the inter-flare interval equals 
half the orbital period when flares from both crossings are observable. At one 
crossing the star moves toward the observer, and half an orbit later it moves away; 
because the emission from the approaching crossing is expected to be stronger than 
that from the receding crossing \citep{huang2025multi,2026JankoviJ,2026HuangS}, 
this scenario naturally produces QPEs with alternating amplitudes and recurrence 
times, as observed in GSN~069, RXJ1301.9+2747, eRO-QPE2, and eRO-QPE4 
\citep{miniutti2019nine,giustini2020x,arcodia2021x,arcodia2024more}. If the far-side 
emission is too faint to detect \citep{huang2025multi}, the inter-flare interval 
instead equals the full orbital period. For a circular orbit around a 
$M_\mathrm{BH} \sim 10^6\,M_\odot$ black hole, the orbital period is $8.6\,\mathrm{hr}$ 
at $R = 100\,r_g$ and $5.3\,\mathrm{days}$ at $R = 600\,r_g$. The companion is likely 
on an extreme mass-ratio inspiral (EMRI) orbit, delivered by gravitational inspiral, 
star--disk interactions, few-body encounters, or in-situ formation in an AGN disk 
\citep[e.g.][]{sari2019tidal,Wang2024,zhou2024probing,Linial2024coupled,jiang2025embers,
naoz2025triples,rom2024dynamics}. For the companion's orbit to intersect the disk, the 
recurrence time also constrains the disk extent, informing models of the disk's origin 
and evolution 
\citep[e.g.][]{linial2023emri+,zhou2024probing,chakraborty2024testing,mummery2025collisions}. 
Secular timing variations have been observed in several systems 
\citep{giustini2020x,Arcodia2022ero1,HernandezGarcia2025,hernandez2025nicer} and 
may constrain disk or orbital precession 
\citep{franchini2023quasi,chakraborty2024testing,zhou2024probing,xian2025secular,
chakraborty2025prospects}.

The flare amplitude can further constrain the perturber and disk properties. 
\citet{linial2023emri+} showed that the total QPE luminosity is consistent with the 
collisional energy between a stellar-sized object and a thin, sub-Eddington accretion 
disk. The interaction is geometric rather than gravitational since the Bondi--Hoyle radius 
$r_\mathrm{Bondi} \sim 2GM_*/v^2 \approx 4.2\times10^{-4}\,R_\odot\,(M_*/M_\odot)
(v_*/0.1c)^{-2}$ is several orders of magnitude smaller than the stellar radius,  or in other words, 
the stellar speed far exceeds the escape velocity from the stellar surface. 
\citet{linial2023emri+} argued that the disk thickness is comparable to the stellar 
radius, and the collision ejects disk material at the impact site; as the ejecta expand 
and cool, the evolving photosphere determines the QPE emission properties. The spectral 
evolution of this ejecta was later modeled by \citet{vurm2025radiation} using 
time-dependent, one-dimensional Monte Carlo radiation-hydrodynamic simulations, 
finding that photon production and reprocessing reproduce both the observed soft X-ray 
luminosity and the characteristic spectral hardening during flares. \citet{huang2025multi} 
studied the broadband SED evolution using two-dimensional, multi-group 
radiation-hydrodynamic simulations and found that the emission properties depend 
sensitively on the assumed opacity, with bound-free opacity promoting photon production 
relative to free-free opacity. Their simulations reproduce the observed soft X-ray 
luminosities and SED evolution. However, the ejecta emission mechanism is short-lived, 
lasting only minutes, and may not account for the longer-duration QPE events without 
additional emission mechanisms. Similar light curve morphologies were found by 
\citet{jankovivc2026radiation}, who studied the collision using frequency-integrated 
radiation hydrodynamics.

The fluid dynamics of the star--disk collision are beginning to be understood. 
\citet{2026HuangS} simulated the interaction between a solid sphere and a thin disk 
using the immersed solid boundary method within the \texttt{Athena++} framework, 
finding that the bow shock ahead of the sphere is strongly compressed upon disk entry. 
Numerically resolving the shock stand-off distance is crucial for capturing the ejecta 
mass and energetics, since most of the collision energy is stored within this compressed 
shock. When the star exits the thin disk from the high-density midplane into the 
low-density corona, the stored shock energy releases rapidly, driving an accelerating, 
detached shock that breaks out and pushes disk material off the disk surface. If instead 
the star has an extended envelope --- for example, after repeated collisions --- the 
envelope may be stripped by the ram pressure \citep{yao2025star}, and its interaction 
with the disk may itself power flares \citep{linial2025qpes}.

In this work, we follow \citet{2026HuangS} and treat the star as a solid sphere, but
focusing on shock propagation through a more realistic AGN disk structure drawn from 
the radiation-MHD simulations of \citet{Jiang2019}, rather than an idealized thin disk. 
The standard Shakura-Sunyaev thin disk model is primarily a vertically integrated model and does not uniquely specify the detailed vertical density structure. 
The inner regions of AGN accretion disks are typically radiation-pressure dominated. In this regime, vertical hydrostatic balance is largely set by the balance between radiation and gravity,
$\kappa_{\rm es} F/c \simeq \Omega^2 z$,
and therefore does not by itself determine the vertical density profile. Instead, the density structure depends sensitively on how the turbulent dissipation and radiative flux are distributed with height (e.g., \citealt{Blaes2011}).
Radiation-MHD simulations of MRI turbulence show that magnetic fields amplified near the disk midplane buoyantly rise toward the disk surface, producing substantial magnetic support and dissipation at large heights. As a result, the vertical density profile generally consists of a relatively narrow dense core surrounded by an extended, more slowly declining atmosphere. The resulting density scale height can therefore be substantially larger than the simple estimate (H$\sim c_s/\Omega$) based on the midplane sound speed. Such vertically extended structures were first studied in detail using local radiation-MHD shearing-box simulations \citep{Blaes2007,Blaes2011,Jiang2013,Jiang2016} and have subsequently also been found in global simulations \citep{Jiang2019,Jiang2025}. More generically, even without radiation pressure support, these vertically extended disks have also been observed in gas pressure dominated MHD disks due to magnetic support (e.g. \citealt{ZhuStone2018}). 

We aim to address two prominent observed trends: the flare duration is $\sim$10--20\% 
of the recurrence period, and longer-period QPEs tend to be more luminous. We also 
consider oblique star--disk collisions across a range of incident angles, as expected 
in realistic EMRI geometries. Due to the relative velocity between the star and the 
disk (e.g., \citealt{Wang2024}), even a star on a circular orbit with its orbital plane 
perpendicular to the disk plane encounters the disk obliquely at $\sim$$45^\circ$ 
owing to the disk's Keplerian motion. The strictly perpendicular impact velocity 
assumed in previous studies 
\citep{yao2025star,huang2025multi,jankovivc2026radiation} arises only when the 
star co-rotates with the disk, which simultaneously implies zero vertical collision 
velocity --- an unphysical limiting case. \citet{2026HuangS} showed that for oblique 
collisions the bow shock expands laterally and breaks out from both the upper and lower 
disk surfaces, producing more symmetric mass and energy ejection.

The remainder of this paper is organized as follows. We describe the methods and 
simulations in Section~\ref{sec:method}, present the results in 
Section~\ref{sec:results}, and discuss and conclude in Section~\ref{sec:summary}.

\begin{figure*}
    \includegraphics[trim=0cm 8.5cm 0.5cm 0.5cm, clip, width=\textwidth]{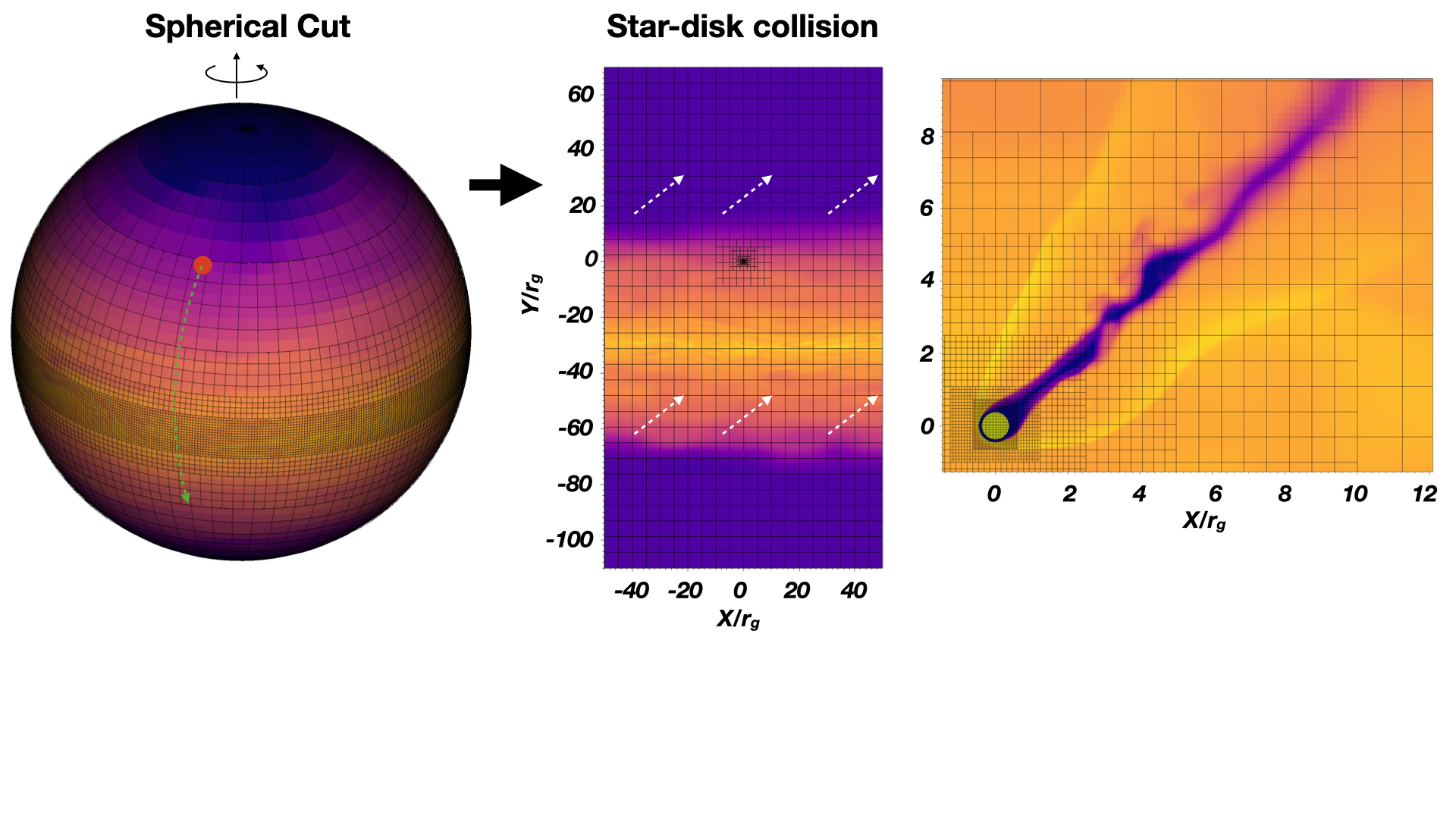}
    \caption{
             Schematic illustration of the procedures adopted in this work. 
(\textit{Left}) A spherical shell at $50\,r_g$ is extracted from the sub-Eddington 
accretion-disk simulation of \citet{Jiang2019} (AGN0.2 run). The red dot and green curve 
mark trajectory of a star crossing the disk. (\textit{Middle}) The 
sphere is unfolded onto a Cartesian $x$--$y$ plane. All hydrodynamic quantities are spatially 
smoothed to suppress grid-scale fluctuations, yielding the initial disk profile used 
in our simulations. The smoothed disk 
structure is used as input for two-dimensional radiation-hydrodynamic (RH) simulations 
of a star--disk collision. The disk is initialized moving toward the star, which is 
treated as a rigid solid sphere (black point). Mesh-refinements with  five additional levels have been adopted towards the star.
 (\textit{Right}) The gas density structure as the star approaches the disk midplane is shown alongside  the simulation grid. Each displayed grid cell represents $8\times8$ computational 
cells.  }
    \phantomsection
    \label{fig:model_cartoon}
\end{figure*}

\section{Methods and Simulations}\label{sec:method}
We carry out two-dimensional radiation-hydrodynamic simulations of the interaction 
between a solid sphere and a realistic circumnuclear disk using the multi-group implicit radiation 
hydrodynamics module \citep{Jiang2021,Jiang2022} in \texttt{Athena++} \citep{stone2020athena++}.

The circumnuclear disk structure is taken from the sub-Eddington accretion-disk 
simulation AGN0.2 of \citet{Jiang2019}, which assumes a central black hole mass 
of $5\times10^8\,M_\odot$. To rescale the disk to a $10^6\,M_\odot$ black hole, 
we proceed as follows. We keep the temperature and velocity units unchanged 
($T_0 = 2\times10^5\,\mathrm{K}$ and $c_{s,0} = 5.26\times10^6\,\mathrm{cm\,s^{-1}}$), 
increase the density and pressure units by a factor of 500, and decrease the length 
and time units by the same factor. Under this rescaling, the mass accretion rate unit 
decreases by a factor of 500, so that the mass accretion rate is still $20\%$ of the Eddington value.

However, the temperature must be recalculated to maintain hydrostatic equilibrium 
for the new black hole mass. We first compute the total pressure (radiation plus 
thermal) from the AGN0.2 simulation,
\begin{equation}
    P_\mathrm{old,cgs} = \frac{a T_\mathrm{old,cgs}^4}{3} + 
    \frac{\rho_\mathrm{old,cgs} k_B T_\mathrm{old,cgs}}{\mu m_H},
\end{equation}
scale it by a factor of 500 for the hydrostatic equilibrium (keeping the fixed $P/\rho$ at any $r_g$),
\begin{equation}
    P_\mathrm{new,cgs} = 500\, P_\mathrm{old,cgs},
\end{equation}
and solve for the new temperature $T_\mathrm{new}$ from
\begin{equation}
    P_\mathrm{new,cgs} = \frac{a T_\mathrm{new,cgs}^4}{3} + 
    \frac{\rho_\mathrm{new,cgs} k_B T_\mathrm{new,cgs}}{\mu m_H}.
\end{equation}
In the dimensionless unit system of \citet{Jiang2019}, this last equation takes the form
\begin{equation}
    \frac{\mathbb{P}_\mathrm{new}\, T_\mathrm{new}^4}{3} + \rho_\mathrm{new} T_\mathrm{new} 
    = \frac{\mathbb{P}_\mathrm{old}\, T_\mathrm{old}^4}{3} + \rho_\mathrm{old} T_\mathrm{old},\label{eq:Pnew}
\end{equation}
where $\rho_{new}=\rho_{old}$, and $\mathbb{P} \equiv a T_0^4 / P_0$ is the radiation pressure constant in code units. Note that if the dimensionless density $\rho_\mathrm{new}$ is increased by some factor relative to $\rho_\mathrm{old}$, the left-hand side of Equation \ref{eq:Pnew} must also be multiplied by 
the same factor to maintain the hydrostatic balance.

After rescaling, we extract a spherical shell at $r_0 = 50\,r_g$ and map all quantities within $\theta \times \phi = [-1.5, 1.5] \times [0, 2\pi]$ onto a Cartesian $X$--$Y$ plane, with $X = r_0\phi$ and $Y = r_0\theta$ (Figure \ref{fig:model_cartoon}). The data are interpolated onto a uniform $512\times512$ grid. Since this resolution exceeds that  of the original AMR data from \citet{Jiang2019}, all hydrodynamic quantities are  spatially smoothed at each refinement level to suppress grid-scale fluctuations; 
the half-width at half-maximum (HWHM) of the smoothing kernel is $\sim$2--3 local grid cells at every SMR level. Although magnetic fields are not included in our 
star--disk collision simulations, we perform divergence cleaning on the mapped Cartesian grid to enforce $\nabla\cdot\mathbf{B} = 0$ for future MHD simulations. Note that, for a star orbiting a $10^6\,M_{\odot}$ black hole at $50\,r_g$, the Keplerian period is ${\sim}3$ hours, at the lower end of observed QPE recurrence times. Moreover, $50\,r_g$ is close to the tidal disruption radius of a solar-mass star around a black hole of this mass, so a disk-crossing radius of $100\,r_g$ would be more appropriate. However, radiation-MHD disk simulations are numerically expensive, and those of \citet{Jiang2019} only marginally reach steady state at $50\,r_g$. Nevertheless, we expect the general physical processes and results to remain robust for a star crossing the disk at $100\,r_g$.

Finally, this background disk structure is interpolated  into two-dimensional radiation hydrodynamic simulation with a hard sphere  \citep{2026HuangS} at the origin that represents a star. The X-Y domain covering 50$r_g$$\times$90$r_g$ has 640$\times$1024 grid cells. Periodic boundary condition is applied at the X-direction, while outflow boundary condition is applied at the Y-direction. Five levels of mesh-refinement are applied towards the origin, with the finest level extending 1.3$R_{\odot}$ in both X and Y directions. 
The star is initially placed at 15 $r_g$ away from the disk midplane. 
The star is resolved by ${\sim}200$ grid cells per stellar diameter, more than twice the resolution required in \citet{2026HuangS}. Furthermore, because the disk vertical density profile in this work is much smoother, the shock standoff distance is not compressed as strongly as in \citet{2026HuangS}, making the resolution requirement even less stringent here.

Since the simulation is performed in the star's rest frame, the fluid is initialized 
moving toward the star. In practice, we subtract the star's orbital velocity 
$\boldsymbol{v}_\mathrm{star}$ from the disk's inertial-frame velocity throughout the entire 
computational domain to model the star--disk collision in the star's rest frame. 
The vertical gravitational acceleration from the central black hole, 
$g_z = -\Omega_K^2\,z$, is applied throughout the domain.

The star is modeled using an immersed solid boundary condition \citep{CHOUNG2021110198,2023Moran,2026HuangS} to divert the gaseous 
flow around it. For the radiation field, a high-density absorbing sphere of 
$\rho = 5\times10^{-4}\,\mathrm{g\,cm^{-3}}$ and $T = 6000\,\mathrm{K}$ is placed 
within $0.8\,R_\odot$, and a uniform electron-scattering opacity is applied within 
$0.85\,R_\odot$ to scatter away most of the radiation. The total electron-scattering 
optical depth across the stellar region is $2\times10^7$. Although both the density 
and optical depth are three orders of magnitude below their solar values, they greatly 
exceed those of the background disk. The buffer region between $0.85$ and $1\,R_\odot$ 
is assigned zero opacity to avoid numerical complications arising from the strong 
radiation field and high optical depth at the stellar surface. In reality, this is the 
region where the star's outer envelope is subject to ram-pressure stripping, and the 
detailed treatment of this interface may be physically important. We defer a more 
complete treatment to future work.

We carry out six simulations in total: a reference run without a stellar crossing, 
four runs with the star moving at incident angles of $90^\circ$, $112^\circ$, 
$135^\circ$, and $157^\circ$ in the disk's co-rotating frame{  , and one further run
repeating the $135^\circ$ collision in a disk whose density, and hence surface density,
is increased everywhere by a factor of $100$, used in Section~\ref{sec:results} to test
how the flare scales with the disk surface density}. Although the 
perpendicular ($90^\circ$) geometry has been widely adopted in previous work and also been carried out here for comparison, it is 
not physically realistic: it requires the star and disk to share the same orbital 
velocity, which only occurs when the star lies in the disk plane, leaving no relative 
vertical motion between the star and the disk. The remaining three cases ($112^\circ$, 
$135^\circ$, and $157^\circ$) correspond to configurations in which the star's 
velocity makes an angle of $45^\circ$, $90^\circ$, and $135^\circ$ with respect to 
the disk's Keplerian velocity. For the $90^\circ$ case, we adopt an impact velocity 
of $\sqrt{2}\,v_K$. For the other cases, the relative velocity is computed from the 
orbital geometry; consequently, the $157^\circ$ case has the highest disk-penetrating 
velocity in the disk's co-rotating frame.

Our RHD simulations have adopted ten frequency groups from $h\nu=10^{-3}$ keV to $h\nu=4$ keV using TOPS Opacity database \citep{Colgan2016}, as in \cite{huang2025multi}. 

\begin{figure}[htb!]
    \includegraphics[width=\columnwidth]{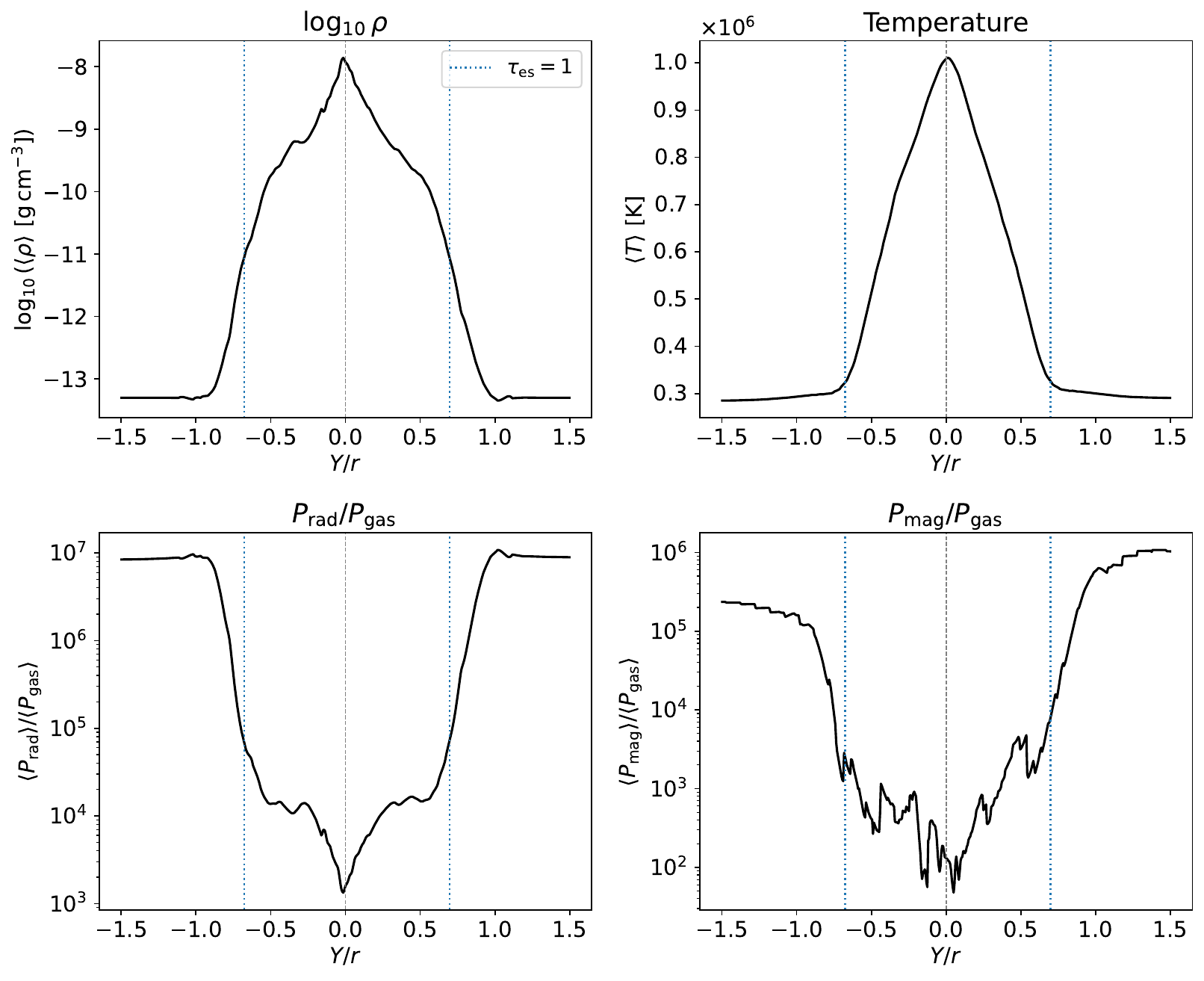}
    \caption{
             Vertical profiles of the disk density $\rho$, gas temperature $T$, 
radiation-to-gas pressure ratio $P_\mathrm{rad}/P_\mathrm{gas}$, and 
magnetic-to-gas pressure ratio $P_\mathrm{mag}/P_\mathrm{gas}$, after scaling 
to a $20\%\,\dot{M}_\mathrm{Edd}$ accretion rate disk around a $10^6\,M_\odot$ black 
hole. Angle brackets denote azimuthal averages. The vertical dotted blue lines 
mark the locations where the electron-scattering optical depth integrated from the 
disk surface reaches $\tau_\mathrm{es} = 1$.
    }
    \label{fig:2D_mach}
\end{figure}

\section{Simulations and Results} \label{sec:results}
The initial disk vertical structure is shown in Figure~\ref{fig:2D_mach}. A notable 
feature is the extended disk atmosphere, which pushes the electron-scattering 
photosphere to $z/r \sim 0.7$. The gas temperature at the $\tau_\mathrm{es} = 1$ 
surface is $\sim$$10^5\,\mathrm{K}$, rising to $\sim$$10^6\,\mathrm{K}$ at the 
midplane. Throughout the disk, radiation pressure dominates over magnetic pressure, 
which in turn exceeds the gas thermal pressure. Even at the midplane, the radiation 
pressure exceeds the gas pressure by three orders of magnitude, confirming that the 
disk is strongly radiation-pressure dominated. The azimuthally averaged surface
density at this radius is $\Sigma \approx 1.7\times10^{4}\,\mathrm{g\,cm^{-2}}$, varying
by ${\sim}\pm25\%$ with azimuth.

\begin{figure*}[htb!]
    \includegraphics[width=\textwidth]{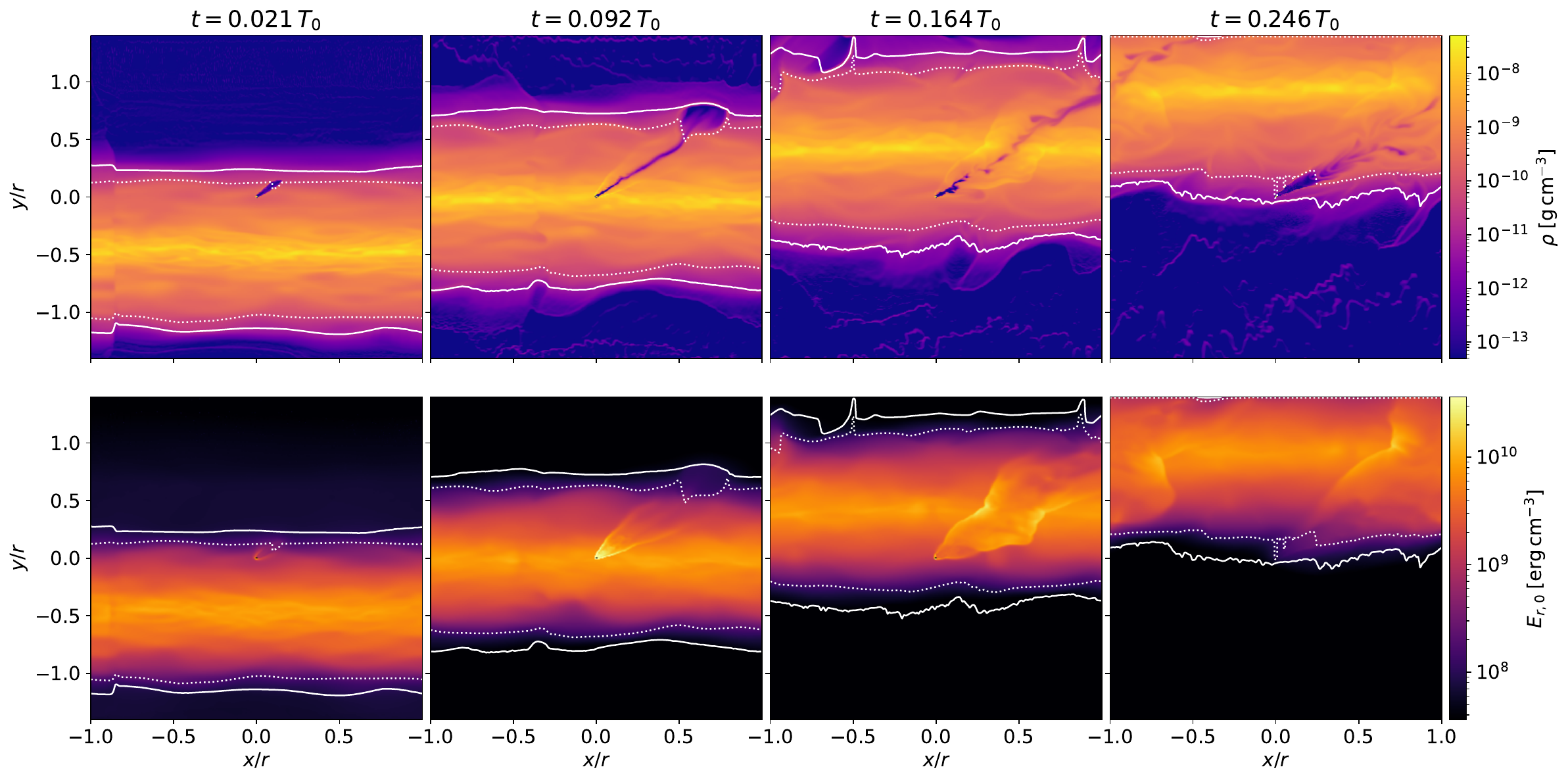}
    \caption{
             Snapshots of the gas density $\rho$ (top row) and radiation energy density 
$E_{r,0}$ (bottom row) during the passage of the solid sphere through the AGN disk, 
at four successive times $t = 0.021,\, 0.092,\, 0.164,$ and $0.246\,T_0$ (left to 
right). The white solid and dotted contours mark the surface where the electron-scattering optical 
depth integrated from above reaches $\tau_\mathrm{es} = 1$ and 7.
    }
    \label{fig:2Dplot}
\end{figure*}

The star-disk crossing for the $135^\circ$ case is illustrated in 
Figure~\ref{fig:2Dplot}, showing snapshots of the gas density $\rho$ (top row) and 
radiation energy density $E_{r,0}$ (bottom row) at four successive times. The star 
leaves a low-density wake behind it; this density void is gradually erased by 
turbulence driven by the disk and the bow shock. Since the star's Keplerian speed is 
$0.14\,c$, corresponding to Mach $\sim$360 for gas at $10^6\,\mathrm{K}$, the star's motion 
is highly supersonic and drives a strong bow shock. In the star's rest frame, the 
kinetic energy of the incoming gas is thermalized at the shock. The bulk of this 
thermal energy is generated and deposited in the post-shock region when the star passes through the disk midplane, where the 
density is highest (second panel from left). After the star crosses the midplane, the 
shock continues to expand but its morphology is progressively shaped by the local disk 
stratification (third panel from left). The bow shock breaks out through the $\tau_\mathrm{es}\sim c/v\sim 7$
surface (dotted white contours, rightmost panel) at approximately the same time as the star itself exits this surface, although the shock breakout location can be displaced from the star's position. We approximate the shock speed as the star's orbital speed. The observable X-ray flare emerges only after this breakout. Different parts of the shock break out at different times, and each breakout region continues to radiate afterwards \citep{Nakar2010}. This picture is similar to the model of \citet{tagawa2023flares}; however, owing to the thick disk structure in our simulations, such extended bow-shock breakout can occur even when the star is not on a low-inclination orbit (e.g., for a nearly vertical crossing orbit).

\begin{figure*}[htb!]
    \centering
    \includegraphics[width=1.0\textwidth]{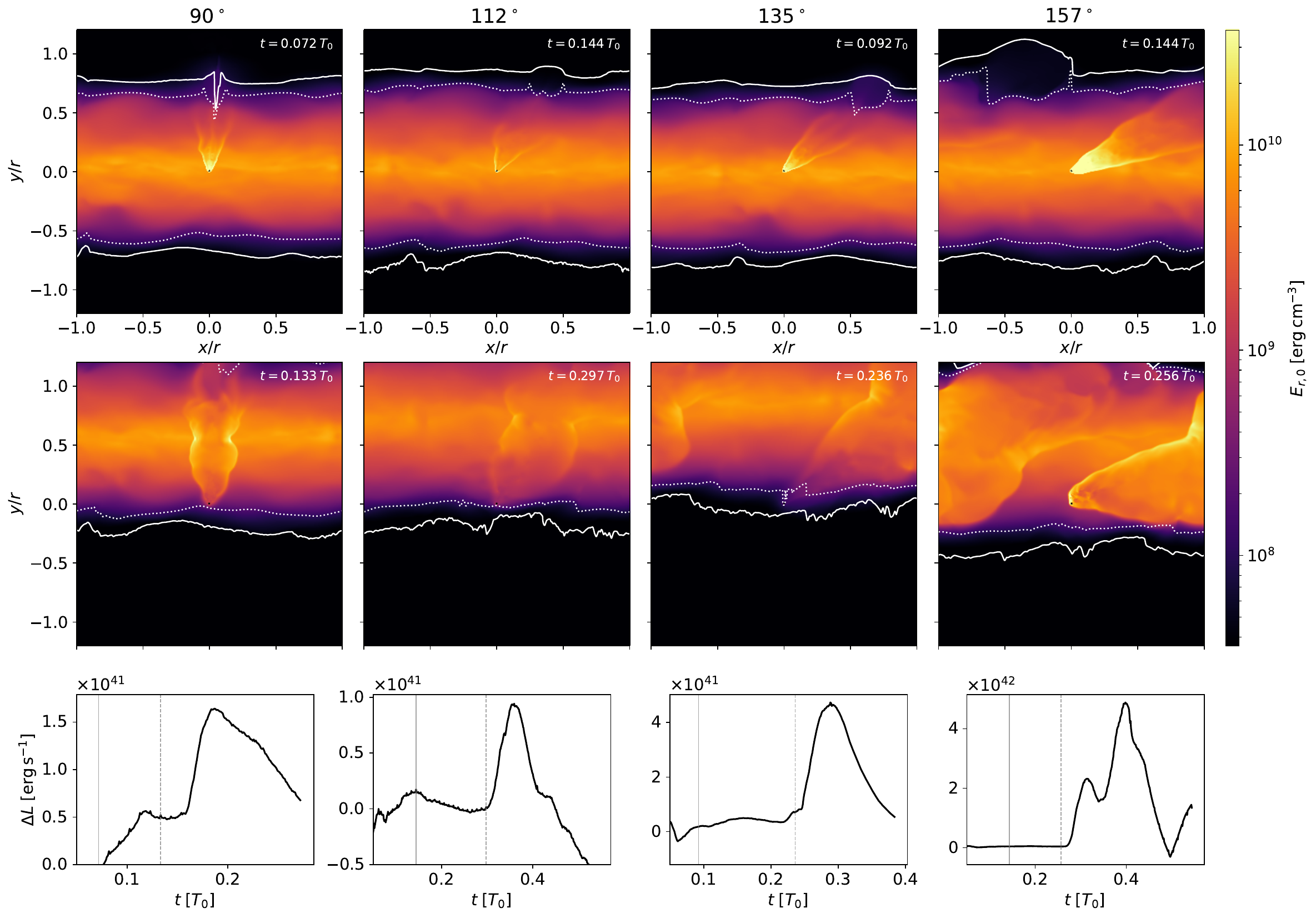}
    \caption{Snapshots of the radiation energy density $E_{r,0}$ at two moments: when 
the star is at the disk midplane (top row) and when the star is about to exit the 
disk (middle row), for four different stellar incident angles $\theta = 90^\circ,\, 
112^\circ,\, 135^\circ,$ and $157^\circ$ in the disk co-orbiting frame (left to 
right). The white solid and dotted contours mark the surface where the electron-scattering optical 
depth integrated from above reaches $\tau_\mathrm{es} = 1$ and 7. For the most oblique 
case ($\theta = 157^\circ$), the bow shock breaks out through the disk surface before 
the star exits, so the middle snapshot is instead taken at the moment of first shock 
breakout. The bottom panels show the corresponding excess luminosity $\Delta L$ as a 
function of time; vertical dashed lines indicate the times of the snapshots shown above.}
    \label{fig:SD_2D_resolu}
\end{figure*}

Figure~\ref{fig:SD_2D_resolu} shows snapshots of the radiation energy density $E_{r,0}$ 
(top and middle rows) and the corresponding excess light curves (bottom row) for all 
four simulations with different stellar incident angles. The top row shows the moment 
when the star is at the disk midplane, and the middle row shows the moment when the 
star is about to exit the disk. For more inclined orbits (excluding the unphysical 
$90^\circ$ case), the relative speed between the star and the disk material is higher, 
producing a stronger bow shock and a more energetic post-shock region.

The excess luminosity $\Delta L$ shown in the bottom panels is measured at 
$y/r = -1.2$ below the disk midplane and is computed by subtracting the 
frequency-integrated flux of the reference (no-star) simulation from that of each 
collision run. This subtraction is necessary because our 2D setup undergoes an initial 
relaxation stage --- arising from the unit rescaling and the absence of driving MHD 
turbulence --- that itself produces a spurious flare. Although the disk density changes 
little during relaxation and does not affect the star--disk interaction, this artifact 
can obscure the shock-breakout flare, particularly for the $90^\circ$ and $112^\circ$ 
cases where the collision flare is weaker. To convert the 2D flux to an emergent 
luminosity, we assume the slice has an effective thickness of $2\,R_\odot$, which 
approximates the cross-sectional impact area of a star in a 3D disk.

The most important result is that the flare duration is $\sim$10--20\% of the stellar 
orbital period $T_0$. This arises naturally from the geometry of shock breakout: luminosity  
increment is produced when the hot post-shock region emerges through the 
$\tau_\mathrm{es} \sim c/v\sim 7$ surface. The oblique shock front spans a length scale 
comparable to the disk thickness $H_\mathrm{es}$ that is the distance from the midplane to the upper or lower 
photosphere, and propagates at approximately the stellar velocity $v_K$. The 
resulting breakout timescale is $\sim H_\mathrm{es}/v_K$; since $H_\mathrm{es} 
\approx 0.7\,r$, this corresponds to $\sim$11\% of the orbital period, consistent 
with the observed QPE duty cycles. Compared to previous models, one essential difference is that the flare duration is set by the dynamical time for the shock to cross the vertically extended disk, rather than by the photon diffusion time. Consequently, the ${\sim}10\%$ duty cycle is largely determined by the disk vertical structure, given that the shock velocity is close to Keplerian. The duration is
correspondingly insensitive to the impact angle: across our four geometries it stays
within 10--20\% of $T_0$, a factor of two, while the peak luminosity varies by a factor
of $>10$. Changing the angle alters which part of the approximately parabolic shock
front reaches the breakout surface first --- for the $157^\circ$ case the trailing end
emerges before the star itself exits --- but not the vertical distance the shock must
traverse.

Since the shock must propagate through the optically thick disk atmosphere before 
breaking out, the rise timescale is a moderate fraction of the total flare duration, 
in contrast to the prompt emission produced by ejecta in star--thin disk collision 
models \citep{linial2023emri+,huang2025multi,vurm2025radiation}. Cooling emission from the expanding ejecta, which powers the flare in the
thin-disk picture of \citet{linial2023emri+}, is included self-consistently in our
multi-group calculation, but it does not contribute much due to the geometric reason. When the
disk thickness is comparable to the stellar radius, the shock exits into the low-density
corona while still carrying most of the collision energy (the successful breakout condition is given in Equation  7 on \citealt{2026HuangS}); the shocked gas then expands
essentially freely, and the cooling radiation from its receding photosphere is the
flare. In our disk the electron-scattering photosphere lies at $z/r \approx 0.7$, so the
shock must instead traverse $H_\mathrm{es}$, some
${\sim}40$ stellar diameters of optically thick material, before any radiation escapes.
There is no prompt release into vacuum: the post-shock gas remains embedded, expands
against the overlying disk material, and shares its energy with the mass it sweeps up as
it climbs. The reservoir available for free expansion is correspondingly small. Only the
column above the breakout surface, $\tau_\mathrm{es} \sim c/v \sim 7$, can expand
ballistically, and that column is ${\sim}22\,\mathrm{g\,cm^{-2}}$, or ${\sim}0.2\%$ of
the disk half-column, so the
thin outer skin that does expand freely carries far too little energy to leave a
distinct signature. The $157^\circ$ case exhibits two distinct luminosity peaks. Careful 
examination of the simulation reveals that this is likely due to the 
limited azimuthal extent of our 2D domain: the star spends an extended period 
traversing the thick disk and re-encounters its own previously driven shock via the 
periodic $X$-boundary. In a realistic 3D disk, such self-interaction would only occur 
for a highly coplanar star on a retrograde orbit.

The flare amplitude is directly related to the impact energy. Measuring the emergent 
flux from $y/r=-1.2$ below the midplane, the total flare energy amounts to $\sim$23\% of the 
impact kinetic energy $E_\mathrm{imp} = \frac{1}{2}\Sigma/{\rm sin(i)} \cdot (2R_\odot)^2 \cdot 
v_\mathrm{rel}^2$ for the $157^\circ$ case. A comparable amount of energy likely 
emerges from the upper disk surface. The remaining energy is deposited into the disk 
and the star; in reality the fraction absorbed by the star is highly uncertain and depends on the stellar structure, which deserves careful studies in future.

\begin{figure*}[htb!]
    \includegraphics[width=1.0\textwidth]{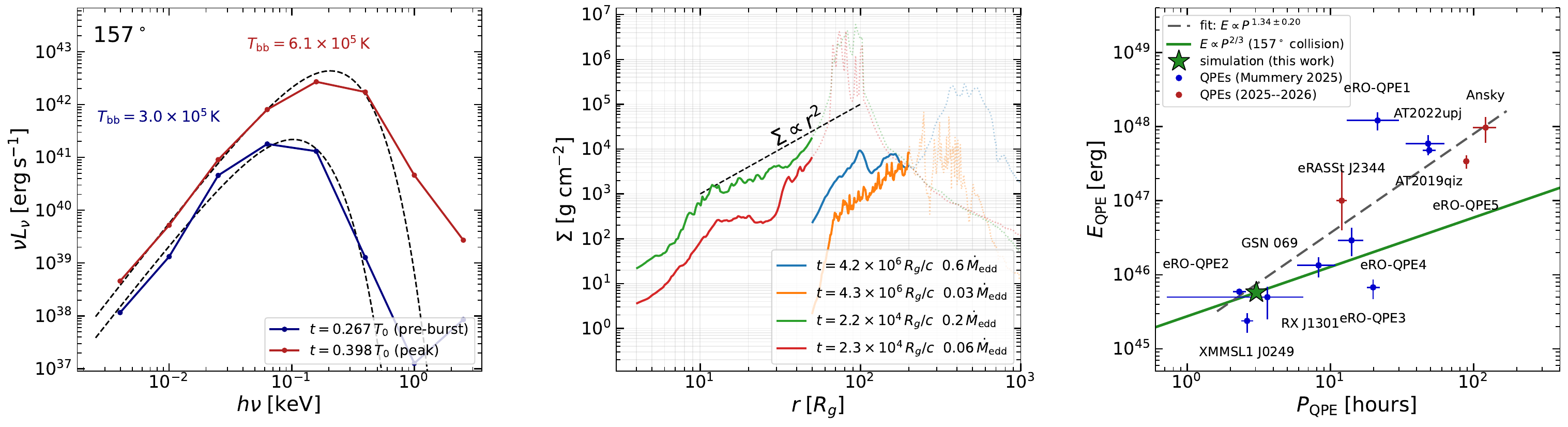}
    \caption{
             (\textit{Left}) Spectral energy distributions of the  luminosity
$\nu L_\nu$ before the outburst ($t = 0.267\,T_0$, pre-burst) and at
peak ($t = 0.398\,T_0$) for the $\theta = 157^\circ$ case. Dashed black curves
show best-fitting blackbody spectra, at $T_\mathrm{bb} = 3.0\times10^5\,\mathrm{K}$
and $T_\mathrm{bb} = 6.1\times10^5\,\mathrm{K}$, respectively. (\textit{Middle}) Disk
surface density $\Sigma$ as a function of radius $r$ at the end of the simulations
for four runs: AGN0.2 (green) and AGN0.07 (red) from \citet{Jiang2019}, and AGNUVB0.6 (blue) and
AGNUVB0.03 (orange) from \citet{Jiang2025}. The dashed black line indicates the
$\Sigma \propto r^2$ scaling. Dotted portions of each curve denote radii
significantly affected by the torus initialization.
{  (\textit{Right}) Radiated energy per eruption $E_\mathrm{QPE}$ against recurrence
time $P_\mathrm{QPE}$ for the observed QPE population: blue points are the
compilation of \citet{mummery2025collisions}, red points are sources reported
since. The green star is the flare energy measured in our $157^\circ$ simulation,
emerging through one disk surface, plotted at the simulated orbital period
$T_0 = 3.04\,\mathrm{hr}$, and the green
line extends it as $E \propto P^{2/3}$ (Equation~\ref{eq:eimp}) with no free
parameters. The grey dashed line is a power-law fit to the observed points.}
    }    \label{fig:mej_2D_resolu}
\end{figure*}

The SEDs before and during the outburst for the $157^\circ$ case are shown in 
Figure~\ref{fig:mej_2D_resolu}. As the luminosity rises, the spectrum hardens, with the 
characteristic blackbody temperature increasing from $\sim$$3\times10^5\,\mathrm{K}$ 
to $\sim$$6\times10^5\,\mathrm{K}$ ($\sim$25\,eV to $\sim$50\,eV). This spectral 
hardening is qualitatively consistent with observations \citep{arcodia2024more,
arcodia2025srg}, and in our model it arises naturally from the shock breakout: the 
post-shock region is hotter than the ambient disk photosphere, so as the shock emerges 
through the $\tau_\mathrm{es} \sim c/v$ surface, the observed emission temperature at $\tau_\mathrm{es}\sim 1$
increases.

\begin{figure*}[htb!]
    \includegraphics[width=1.0\textwidth]{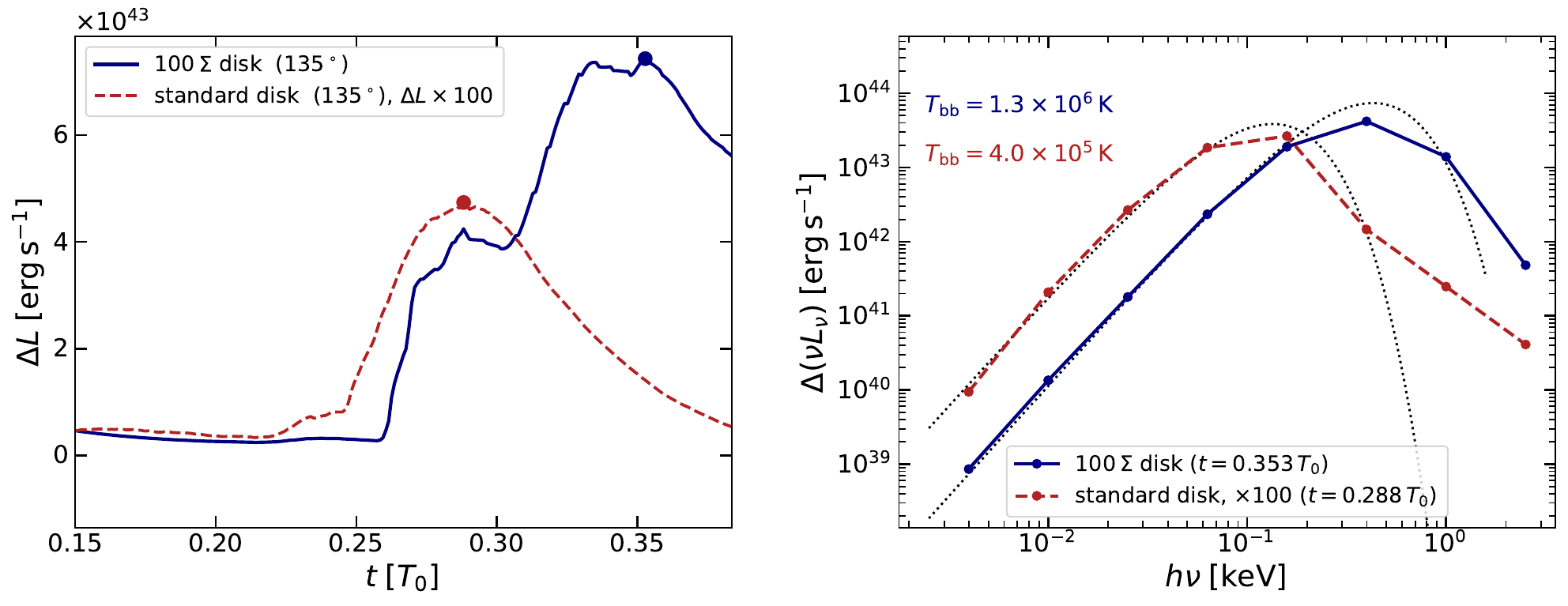}
    \caption{
             Effect of the disk surface density on the flare, for the
$\theta = 135^\circ$ collision. (\textit{Left}) Excess luminosity $\Delta L$,
measured at $y/r = -1.2$ below the midplane after subtracting the reference run
without a star, for a disk with $100\times$ the fiducial surface density
($100\,\Sigma$, blue solid) and for the standard disk with its $\Delta L$
multiplied by $100$ (red dashed). The time axis begins at $t = 0.15\,T_0$ to
exclude the initial relaxation transient. Filled circles mark the light-curve
peaks. (\textit{Right}) Excess spectra $\Delta(\nu L_\nu)$ at those two peaks,
with best-fitting single-temperature blackbodies (dotted) at
$T_\mathrm{bb} = 1.3\times10^6\,\mathrm{K}$ ($109\,\mathrm{eV}$) and
$T_\mathrm{bb} = 4.0\times10^5\,\mathrm{K}$ ($34\,\mathrm{eV}$), respectively.
    }    \label{fig:100times}
\end{figure*}

The scaling of the flare energy with disk surface density can be tested
directly. We repeated the $135^\circ$ collision in a disk whose density, and
hence surface density, is increased by a factor of $100$. Since
$E_\mathrm{imp} \propto \Sigma$, the impact energy is $100$ times larger, and
the emitted luminosity should scale in the same way. Figure~\ref{fig:100times}
shows that it very nearly does: multiplying the standard run's excess
luminosity by $100$ brings the two light curves into agreement in both
amplitude and overall shape, confirming that the flare energy is set by the
column swept up by the star.

The agreement is not exact, and the difference is instructive. The peak of the
denser run reaches $7.4\times10^{43}\,\mathrm{erg\,s^{-1}}$ against
$4.7\times10^{43}\,\mathrm{erg\,s^{-1}}$ for the rescaled standard run, and it
arrives later, at $t = 0.353\,T_0$ rather than $0.288\,T_0$. A more massive disk
is also more optically thick, so the shock-deposited energy takes longer to
escape. In our setup the radiation does not emerge before the shock has
propagated around the periodic $X$-domain, so the shock front collides with
itself and the same region of the disk is heated twice, producing a single
stronger and later peak. As with the double-peaked light curve of the
$157^\circ$ case discussed above, this behavior depends on the azimuthal
extent of the domain; in a real disk the shock would
instead continue to expand more in the radial and azimuthal directions.

The more energetic flare is also spectrally harder. The best-fitting blackbody
temperature rises from $34\,\mathrm{eV}$ in the standard disk to
$109\,\mathrm{eV}$ in the $100\,\Sigma$ disk, a factor of $3.2$ close to the
factor $100^{1/4} \simeq 3.2$ expected if the luminosity increases by two orders
of magnitude at a roughly unchanged emitting area\footnote{For all our simulations, the effective emitting area is 10-30 $(2R_{\odot})^2$. A more inclined collision is at the higher end of the effective area. }. This brings the flare
temperature into the $\sim$100--200\,eV range characteristic of observed QPEs,
and suggests that the relatively soft emission of our fiducial runs could reflect the
modest surface density of the rescaled AGN0.2 disk rather than a limitation of
the shock-breakout mechanism itself. A real QPE host disk is probably less magnetized than the AGN0.2 disk so that it has a higher surface density with a smaller accretion efficiency $\alpha$.
On the other hand, we note that our simulations do not
include Comptonization. Figure 16 in \cite{huang2025multi} has shown that Compton scattering could harden the peak of the spectrum by a factor of
 2, and thus the temperatures quoted here should be regarded as lower limits.

Another observed trend in QPEs is that more luminous flares tend to occur in 
longer-period systems \citep{nicholl2024quasi,mummery2025collisions}. Although this 
seems counterintuitive --- stars at larger orbital radii have lower Keplerian kinetic 
energy --- we argue that this trend arises naturally once the surface density profile 
of the circumnuclear disk is accounted for. As shown in the right panel of Figure~\ref{fig:mej_2D_resolu}, the four sub-Eddington accretion disk simulations of \citet{Jiang2019} and \citet{Jiang2025} all approximately follow $\Sigma \propto r^2$. The reversal of this trend at larger radii is a consequence of the limited disk evolution time imposed by computational cost: the simulations in \citet{Jiang2025} run considerably longer than those in \citet{Jiang2019} and accordingly extend the relation to larger radii. In a real AGN or TDE disk, this trend may persist to still larger radii, since even the longest simulation duration of $5\times10^{6}\,r_g/c$ corresponds to only ${\sim}9$ months for a $10^6\,M_\odot$ black hole — shorter than the typical delay between a TDE and the onset of QPEs.
Combined with the Keplerian scaling $v_K \propto r^{-1/2}$, the total impact kinetic 
energy is
\begin{equation}
    E_\mathrm{imp} = \frac{1}{2}\Sigma \cdot (2R_\odot)^2 \cdot v_\mathrm{rel}^2 
    \propto r \propto P^{2/3}\,,\label{eq:eimp}
\end{equation}
with the circular Keplerian orbit assumption. 
This scaling is tied to the disk radial structure ($\Sigma\propto r^2$) informed by realistic RMHD simulations \citep{Jiang2019,Jiang2025}, and it
reproduces the sense of the observed luminosity--period trend: more energetic
flares occur in longer-period systems, despite the lower Keplerian kinetic
energy available at larger radii.

{  
The right panel of Figure~\ref{fig:mej_2D_resolu} compares this prediction with
the observed population. We adopt the convention of
\citet{mummery2025collisions}, $E_\mathrm{QPE} \approx L_\mathrm{bol,peak}\,
t_\mathrm{QPE}$, and take his compilation of nine sources (GSN\,069,
RX\,J1301.9+2747, XMMSL1\,J0249, eRO-QPE1--4, AT2019qiz and AT2022upj; blue
points). To these we add three sources reported subsequently (red points):
eRO-QPE5 \citep{arcodia2025srg}, with a published eruption energy of
$(3.4\pm0.7)\times10^{47}\,\mathrm{erg}$; Ansky
\citep{HernandezGarcia2025,hernandez2025nicer}, for which we plot the published
average integrated energy per burst of the 2024 discovery epoch,
$(9.7\pm3.7)\times10^{47}\,\mathrm{erg}$, at a recurrence time of
$436.8\pm80.3\,\mathrm{ks}$ (the eruptions had roughly quadrupled in energy by
2025, which we do not plot); and eRASSt~J2344 \citep{Baldini2026}, for which no
eruption energy has been published and we therefore apply the same convention
ourselves. In this last source the broad eruptions carry a crest of hotter,
much shorter flares, so we use only the epochs free of them, giving a mean
quiescent-subtracted bolometric peak of
$2.5\times10^{43}\,\mathrm{erg\,s^{-1}}$ for the two well-covered eruptions;
combined with the broad-flare FWHM of $1.0\,\mathrm{hr}$ this yields
$E_\mathrm{QPE} \approx 1.0\times10^{47}\,\mathrm{erg}$.
In every case the horizontal bar gives the dispersion in the
measured recurrence times, eruption to eruption, rather than the uncertainty on
any single interval: $\pm0.02\,\mathrm{d}$ about a mean of $3.70\,\mathrm{d}$
for eRO-QPE5, $\pm80.3\,\mathrm{ks}$ about $436.8\,\mathrm{ks}$ for Ansky, and
the quoted $12\pm1\,\mathrm{hr}$ for eRASSt~J2344. The vertical bars for
eRO-QPE5 and Ansky are the uncertainties published with their eruption
energies. That for eRASSt~J2344 is instead a systematic range, since the energy
is ours rather than the authors': its lower end
($4\times10^{46}\,\mathrm{erg}$) takes the weakest eruption's broad-flare peak
with the flare FWHM, and its upper end ($2.6\times10^{47}\,\mathrm{erg}$) takes
the blended single-blackbody peak with the nominal $\sim$2~hr duration, that is,
what one obtains by attributing the narrow-flare crest entirely to the
eruption. The bars on the nine sources drawn from
\citet{mummery2025collisions} are those plotted in his Figure~1.

The green star is the flare energy measured in our $157^\circ$ simulation,
$5.8\times10^{45}\,\mathrm{erg}$ emerging through one disk surface, plotted at
the simulated orbital period $T_0 = 3.04\,\mathrm{hr}$; the green line extends
this single anchor with the predicted slope and so contains no free parameters.

A power-law fit to the twelve observed sources, weighted by the quoted
uncertainties, gives $E_\mathrm{QPE} \propto P_\mathrm{QPE}^{1.34\pm0.20}$,
and is stable to the removal of any individual source. Fitting only
\citeauthor{mummery2025collisions}'s nine sources without weights returns
$1.67\pm0.43$, recovering the $E_\mathrm{QPE} \propto P_\mathrm{QPE}^{5/3}$ he
quotes. The predicted index of $2/3$ is therefore shallower than the observed
correlation, by roughly a factor of two in the exponent. The agreement is
considerably closer, however, for the weaker half of the population. Dividing
the sample at a factor of three from the prediction, the six least energetic
sources ($E_\mathrm{QPE} = 2.4\times10^{45}$--$2.9\times10^{46}\,\mathrm{erg}$;
eRO-QPE2, XMMSL1\,J0249, RX\,J1301, GSN\,069, eRO-QPE3 and eRO-QPE4) scatter
about the predicted line. We
therefore regard the model as accounting for the luminosity--period trend of
the weaker QPEs, and note three possible reasons for the overall disagreements, especially for brighter bursts. First, the normalisation itself
carries a geometric uncertainty of a factor of a few: the observed energies are derived assuming the radiation is 
isotropic, whereas our derived value is the energy radiated through a single
disk face. Second, the real disk's surface density profile may be steeper than $\Sigma\propto r^2$ at larger distances. The slopes in the middle panel of Figure \ref{fig:mej_2D_resolu} only very roughly follow  $r^2$.  The disk's surface density at large distances is also strongly affected by the disk's initial condition. 
Third, stronger bursts may emerge from a more optically thick disk, and shock collisions or heated background disks may lead to brighter and hardened radiation, as in Figure \ref{fig:100times}. 
Testing these requires collision simulations at more than one orbial radius, which we defer to future work.
}

\section{Limitation and conclusion}\label{sec:summary}
Although our simulations provide a viable explanation for several observed QPE trends, 
several limitations should be borne in mind. First, we cannot measure the flux emerging 
from the side of the disk opposite to the stellar motion: because the disk is 
geometrically thick, the photosphere approaches the boundary of the computational domain 
as the midplane-generated shock propagates outward. This is a significant limitation, as 
previous simulations with thin disks find that the forward and 
backward ejecta differ substantially in mass, energy, and luminosity---by up to an order 
of magnitude depending on the obliquity of the collision 
\citep{huang2025multi, 2026HuangS, liu2026quasi, Jankovic2026}---and this 
asymmetry has been invoked both to explain the alternating strong--weak flare amplitudes 
observed in several QPE sources and to suggest that only one flare per orbit may be 
detectable. Whether a comparable asymmetry survives in a realistic, geometrically thick 
disk, where the shocked gas must traverse several scale heights before breakout, remains 
an open question. In future work, we will simulate a moving star within a spatially fixed 
disk, allowing the breakout geometry on both faces of the disk to be captured 
simultaneously. Second, the stellar structure is not treated self-consistently: the star 
is modeled as a solid sphere. In reality, repeated collisions progressively heat and 
inflate the stellar envelope, enhancing the mass stripped per passage 
\citep{yao2025star}, and the stripped debris can fall back onto the disk and generate 
additional collisions and flares \citep{Linial2025}. Tidal heating may further accelerate 
the star's inflation and eventual destruction at the small orbital separations relevant 
to QPEs \citep{Yao2026}. Capturing these processes would require replacing the solid 
boundary with a resolved gaseous envelope, which we defer to future work; how the 
stripping and fallback proceed in a thick disk, where the ram-pressure profile 
encountered by the star differs qualitatively from the thin-disk case, has yet to be 
explored. Third, our simulations are two-dimensional. In 3D, the shocked gas can flow 
around the star in the azimuthal direction, reducing the shock standoff distance 
\citep{2026HuangS}, and the shock front expands and curves in the radial direction, 
modifying the breakout geometry and the emitting area \citep{tagawa2023flares, 
Jankovic2026}. Fourth, we have simulated only a single disk crossing. Over many orbits, 
the cumulative energy deposition---comparable to or exceeding the local viscous heating 
rate---could thicken the disk and alter its vertical structure, thereby modifying 
subsequent collisions; a tall shearing-box simulation encompassing many successive 
crossings would be well suited to capturing this feedback.

Nevertheless, in this work, we carry out two-dimensional, multi-frequency radiation-hydrodynamic 
(RH) simulations of star--disk collisions, adopting a realistic circumnuclear disk 
structure from the sub-Eddington accretion-disk simulations of \citet{Jiang2019,
Jiang2025}. We rescale the disk structure to a $10^6\,M_\odot$ black hole accreting 
at $20\%\,\dot{M}_\mathrm{Edd}$, extract a spherical shell at $50\,r_g$, and map it 
onto a 2D Cartesian domain. The disk is geometrically thick, with an electron-scattering photosphere at 
$z/r \sim 0.7$. The disk is strongly radiation-pressure dominated throughout.

We simulate four stellar incident angles ($90^\circ$, $112^\circ$, $135^\circ$, and 
$157^\circ$) in the disk co-rotating frame, spanning the range from the unphysical 
perpendicular geometry to highly oblique encounters expected in realistic EMRI 
systems. In all cases, the supersonic stellar motion drives a strong bow shock. The 
thick disk atmosphere causes different portions of the shock front to break out 
 at different times, producing prolonged 
thermal emission at the $\tau_{es}\sim1-10$ surface rather than the prompt ejecta flash seen in star--thin disk models. 
The flare duration is set by the shock propagation time across the disk atmosphere, 
$\sim H_\mathrm{es}/v_K \sim 0.11\,T_0$, naturally reproducing the observed 
$\sim$10--20\% duty cycle. More oblique collisions produce stronger and longer flares, 
and the strongest collision case ($157^\circ$ case) exhibits spectral hardening during the rise --- consistent 
with observations --- as the hotter post-shock region emerges through the cooler disk 
photosphere. The total flare energy emerging from one side is $\sim$20\% of the impact kinetic energy 
$E_\mathrm{imp} = \frac{1}{2}\Sigma/{\rm sin(i)}\,(2R_\odot)^2\,v_\mathrm{rel}^2$.

The four sub-Eddington disk simulations of \citet{Jiang2019} and \citet{Jiang2025} 
all exhibit $\Sigma \propto r^2$. Combined with the Keplerian scaling 
$v_K \propto r^{-1/2}$, this gives $E_\mathrm{imp} \propto r \propto P^{2/3}$, 
providing a natural explanation for the observed trend that more luminous QPEs occur 
in longer-period systems, and matching both the normalisation and the slope of
the correlation traced by the weaker QPEs; the most energetic events lie above this
scaling and may indicate different disk conditions at larger radii. Overall, we suggest that a realistic circumnuclear disk 
structure --- in particular its geometric thickness and surface density profile --- 
can simultaneously explain several observed QPE trends. Conversely, QPE observations 
may themselves serve as a novel probe of the circumnuclear disk structure around 
quiescent or low-luminosity SMBHs.

\begin{acknowledgments}
ZZ thanks Andrew Mummery, Brian Metzger, and Itai Linial for discussions. The authors thank the reviewer for a constructive report. Simulations were carried out on Flatiron Institute supercomputers. 
ZZ acknowledges support from the Flatiron Institute visitor program, IAS visitor program, NSF awards 2429732 and 2408207. XH is supported by the Sherman Fairchild Postdoctoral Fellowship at the California Institute of Technology. This research also benefited from interactions that were funded by the Gordon and Betty Moore Foundation through Grant GBMF5076. 
\end{acknowledgments}

\begin{contribution}

This project originated during the sabbatical visit of Z.Z.\ to the Institute for 
Advanced Study (IAS) and the Center for Computational Astrophysics (CCA). Z.Z.\ 
carried out the simulations and analysis with contributions from X.H.\ and Y.J. All authors 
contributed to the final manuscript. This manuscript benefited from language editing and proofreading using ChatGPT (OpenAI 2026) and Claude (Anthropic 2026).


\end{contribution}

%


\software{\texttt{Athena++} \citep{2008ApJS..178..137S,2020ApJS..249....4S}
          }


\bibliography{QPE}

@ARTICLE{Baldini2026,
       author = {{Baldini}, P. and {Rau}, A. and {Merloni}, A. and {Trakhtenbrot}, B. and {Arcodia}, R. and {Giustini}, M. and {Miniutti}, G. and {Brennan}, S.~J. and {Freyberg}, M. and {S{\'a}nchez-S{\'a}ez}, P. and {Grotova}, I. and {Liu}, Z. and {Lian}, T. and {Nandra}, K.},
        title = "{Discovery of crested quasi-periodic eruptions following the most luminous SRG/eROSITA tidal disruption event}",
      journal = {\aap},
         year = 2026,
        month = feb,
       volume = {706},
          eid = {L15},
        pages = {L15},
          doi = {10.1051/0004-6361/202558241},
archivePrefix = {arXiv},
       eprint = {2602.03932},
 primaryClass = {astro-ph.HE},
       adsurl = {https://ui.adsabs.harvard.edu/abs/2026A&A...706L..15B}
}

@ARTICLE{Jiang2013,
       author = {{Jiang}, Yan-Fei and {Stone}, James M. and {Davis}, Shane W.},
        title = "{Saturation of the Magneto-rotational Instability in Strongly Radiation-dominated Accretion Disks}",
      journal = {\apj},
         year = 2013,
        month = apr,
       volume = {767},
       number = {2},
          eid = {148},
        pages = {148},
          doi = {10.1088/0004-637X/767/2/148},
archivePrefix = {arXiv},
       eprint = {1303.1823},
 primaryClass = {astro-ph.HE},
       adsurl = {https://ui.adsabs.harvard.edu/abs/2013ApJ...767..148J}
}

@ARTICLE{Jiang2016,
       author = {{Jiang}, Yan-Fei and {Davis}, Shane W. and {Stone}, James M.},
        title = "{Iron Opacity Bump Changes the Stability and Structure of Accretion Disks in Active Galactic Nuclei}",
      journal = {\apj},
         year = 2016,
        month = aug,
       volume = {827},
       number = {1},
          eid = {10},
        pages = {10},
          doi = {10.3847/0004-637X/827/1/10},
archivePrefix = {arXiv},
       eprint = {1601.06836},
 primaryClass = {astro-ph.HE},
       adsurl = {https://ui.adsabs.harvard.edu/abs/2016ApJ...827...10J}
}

@ARTICLE{Blaes2007,
       author = {{Blaes}, Omer and {Hirose}, Shigenobu and {Krolik}, Julian H.},
        title = "{Surface Structure in an Accretion Disk Annulus with Comparable Radiation and Gas Pressure}",
      journal = {\apj},
         year = 2007,
        month = aug,
       volume = {664},
       number = {2},
        pages = {1057-1071},
          doi = {10.1086/519516},
archivePrefix = {arXiv},
       eprint = {0705.0314},
 primaryClass = {astro-ph},
       adsurl = {https://ui.adsabs.harvard.edu/abs/2007ApJ...664.1057B}
}

@ARTICLE{Blaes2011,
       author = {{Blaes}, Omer and {Krolik}, Julian H. and {Hirose}, Shigenobu and {Shabaltas}, Natalia},
        title = "{Dissipation and Vertical Energy Transport in Radiation-dominated Accretion Disks}",
      journal = {\apj},
         year = 2011,
        month = jun,
       volume = {733},
       number = {2},
          eid = {110},
        pages = {110},
          doi = {10.1088/0004-637X/733/2/110},
archivePrefix = {arXiv},
       eprint = {1103.5052},
 primaryClass = {astro-ph.HE},
       adsurl = {https://ui.adsabs.harvard.edu/abs/2011ApJ...733..110B}
}

@ARTICLE{2026HuangS,
       author = {{Huang}, Shunquan and {Huang}, Xiaoshan and {Zhu}, Zhaohuan and {Martin}, Rebecca G.},
        title = "{Resolving Oblique Star-Disk Collisions in Quasi-Periodic Eruptions: Numerical Requirements and the Importance of Geometry}",
      journal = {arXiv e-prints},
         year = 2026,
        month = apr,
          eid = {arXiv:2604.00953},
        pages = {arXiv:2604.00953},
          doi = {10.48550/arXiv.2604.00953},
archivePrefix = {arXiv},
       eprint = {2604.00953},
 primaryClass = {astro-ph.HE},
       adsurl = {https://ui.adsabs.harvard.edu/abs/2026arXiv260400953H}
}

@ARTICLE{Colgan2016,
       author = {{Colgan}, J. and {Kilcrease}, D.~P. and {Magee}, N.~H. and {Sherrill}, M.~E. and {Abdallah}, Jr., J. and {Hakel}, P. and {Fontes}, C.~J. and {Guzik}, J.~A. and {Mussack}, K.~A.},
        title = "{A New Generation of Los Alamos Opacity Tables}",
      journal = {\apj},
         year = 2016,
        month = feb,
       volume = {817},
       number = {2},
          eid = {116},
        pages = {116},
          doi = {10.3847/0004-637X/817/2/116},
archivePrefix = {arXiv},
       eprint = {1601.01005},
 primaryClass = {astro-ph.SR},
       adsurl = {https://ui.adsabs.harvard.edu/abs/2016ApJ...817..116C}
}

@ARTICLE{2026JankoviJ,
       author = {{Jankovi{\v{c}}}, Taj and {Bonnerot}, Cl{\'e}ment and {Karpov}, Sergey and {Jurca}, Aleksej},
        title = "{Radiation-hydrodynamics of star-disc collisions for quasi-periodic eruptions}",
      journal = {arXiv e-prints},
         year = 2026,
        month = feb,
          eid = {arXiv:2602.02656},
        pages = {arXiv:2602.02656},
          doi = {10.48550/arXiv.2602.02656},
archivePrefix = {arXiv},
       eprint = {2602.02656},
 primaryClass = {astro-ph.HE},
       adsurl = {https://ui.adsabs.harvard.edu/abs/2026arXiv260202656J}
}

@ARTICLE{Jiang2021,
       author = {{Jiang}, Yan-Fei},
        title = "{An Implicit Finite Volume Scheme to Solve the Time-dependent Radiation Transport Equation Based on Discrete Ordinates}",
      journal = {\apjs},
         year = 2021,
        month = apr,
       volume = {253},
       number = {2},
          eid = {49},
        pages = {49},
          doi = {10.3847/1538-4365/abe303},
archivePrefix = {arXiv},
       eprint = {2102.02212},
 primaryClass = {astro-ph.IM},
       adsurl = {https://ui.adsabs.harvard.edu/abs/2021ApJS..253...49J}
}

@ARTICLE{Jiang2022,
       author = {{Jiang}, Yan-Fei},
        title = "{Multigroup Radiation Magnetohydrodynamics Based on Discrete Ordinates including Compton Scattering}",
      journal = {\apjs},
         year = 2022,
        month = nov,
       volume = {263},
       number = {1},
          eid = {4},
        pages = {4},
          doi = {10.3847/1538-4365/ac9231},
archivePrefix = {arXiv},
       eprint = {2209.06240},
 primaryClass = {astro-ph.IM},
       adsurl = {https://ui.adsabs.harvard.edu/abs/2022ApJS..263....4J}
}

@ARTICLE{Jiang2019,
       author = {{Jiang}, Yan-Fei and {Blaes}, Omer and {Stone}, James M. and {Davis}, Shane W.},
        title = "{Global Radiation Magnetohydrodynamic Simulations of sub-Eddington Accretion Disks around Supermassive Black Holes}",
      journal = {\apj},
         year = 2019,
        month = nov,
       volume = {885},
       number = {2},
          eid = {144},
        pages = {144},
          doi = {10.3847/1538-4357/ab4a00},
archivePrefix = {arXiv},
       eprint = {1904.01674},
 primaryClass = {astro-ph.HE},
       adsurl = {https://ui.adsabs.harvard.edu/abs/2019ApJ...885..144J}
}

@ARTICLE{Jiang2025,
       author = {{Jiang}, Yan-Fei and {Blaes}, Omer and {Kaul}, Ish and {Zhang}, Lizhong},
        title = "{Radiation and Magnetic Pressure Support in Accretion Disks Around Supermassive Black Holes and the Physical Origin of the Extreme-ultraviolet to Soft X-Ray Spectrum}",
      journal = {\apj},
         year = 2025,
        month = jul,
       volume = {988},
       number = {1},
          eid = {43},
        pages = {43},
          doi = {10.3847/1538-4357/addecb},
archivePrefix = {arXiv},
       eprint = {2505.09671},
 primaryClass = {astro-ph.HE},
       adsurl = {https://ui.adsabs.harvard.edu/abs/2025ApJ...988...43J}
}

@ARTICLE{2008ApJS..178..137S,
       author = {{Stone}, James M. and {Gardiner}, Thomas A. and {Teuben}, Peter and {Hawley}, John F. and {Simon}, Jacob B.},
        title = "{Athena: A New Code for Astrophysical MHD}",
      journal = {\apjs},
         year = 2008,
        month = sep,
       volume = {178},
       number = {1},
        pages = {137-177},
          doi = {10.1086/588755},
archivePrefix = {arXiv},
       eprint = {0804.0402},
 primaryClass = {astro-ph},
       adsurl = {https://ui.adsabs.harvard.edu/abs/2008ApJS..178..137S}
}

@ARTICLE{2020ApJS..249....4S,
       author = {{Stone}, James M. and {Tomida}, Kengo and {White}, Christopher J. and {Felker}, Kyle G.},
        title = "{The Athena++ Adaptive Mesh Refinement Framework: Design and Magnetohydrodynamic Solvers}",
      journal = {\apjs},
         year = 2020,
        month = jul,
       volume = {249},
       number = {1},
          eid = {4},
        pages = {4},
          doi = {10.3847/1538-4365/ab929b},
archivePrefix = {arXiv},
       eprint = {2005.06651},
 primaryClass = {astro-ph.IM},
       adsurl = {https://ui.adsabs.harvard.edu/abs/2020ApJS..249....4S}
}

@article{CHOUNG2021110198,
title = {Nonlinear weighting process in ghost-cell immersed boundary methods for compressible flow},
journal = {Journal of Computational Physics},
volume = {433},
pages = {110198},
year = {2021},
issn = {0021-9991},
doi = {https://doi.org/10.1016/j.jcp.2021.110198},
url = {https://www.sciencedirect.com/science/article/pii/S0021999121000930},
author = {Hanahchim Choung and Vignesh Saravanan and Soogab Lee and Haeseong Cho}
}

@inbook{2023Moran,
author = {Moran Ezra and Yoram Kozak},
title = {Development of an Immersed Boundary Method for High-Speed Compressible Flows},
booktitle = {AIAA SCITECH 2023 Forum},
chapter = {},
year = {2023},
pages = {},
doi = {10.2514/6.2023-1401},
URL = {https://arc.aiaa.org/doi/abs/10.2514/6.2023-1401},
eprint = {https://arc.aiaa.org/doi/pdf/10.2514/6.2023-1401}
}

@article{stone2020athena++,
  title={The athena++ adaptive mesh refinement framework: Design and magnetohydrodynamic solvers},
  author={Stone, James M and Tomida, Kengo and White, Christopher J and Felker, Kyle G},
  journal={The Astrophysical Journal Supplement Series},
  volume={249},
  number={1},
  pages={4},
  year={2020},
  publisher={IOP Publishing}
}

@article{linial2023emri+,
  title={EMRI+ TDE= QPE: periodic X-ray flares from star--disk collisions in galactic nuclei},
  author={Linial, Itai and Metzger, Brian D},
  journal={The Astrophysical Journal},
  volume={957},
  number={1},
  pages={34},
  year={2023},
  publisher={IOP Publishing}
}

@article{metzger2022interacting,
  title={Interacting stellar EMRIs as sources of quasi-periodic eruptions in galactic nuclei},
  author={Metzger, Brian D and Stone, Nicholas C and Gilbaum, Shmuel},
  journal={The Astrophysical Journal},
  volume={926},
  number={1},
  pages={101},
  year={2022},
  publisher={IOP Publishing}
}

@ARTICLE{yao2025star,
       author = {{Yao}, Philippe Z. and {Quataert}, Eliot and {Jiang}, Yan-Fei and {Lu}, Wenbin and {White}, Christopher J.},
        title = "{Star‑Disk Collisions: Implications for Quasi-periodic Eruptions and Other Transients near Supermassive Black Holes}",
      journal = {\apj},
         year = 2025,
        month = jan,
       volume = {978},
       number = {1},
          eid = {91},
        pages = {91},
          doi = {10.3847/1538-4357/ad8911},
archivePrefix = {arXiv},
       eprint = {2407.14578},
 primaryClass = {astro-ph.HE},
       adsurl = {https://ui.adsabs.harvard.edu/abs/2025ApJ...978...91Y}
}

@article{tagawa2023flares,
  title={Flares from stars crossing active galactic nucleus discs on low-inclination orbits},
  author={Tagawa, Hiromichi and Haiman, Zolt{\'a}n},
  journal={Monthly Notices of the Royal Astronomical Society},
  volume={526},
  number={1},
  pages={69--79},
  year={2023},
  publisher={Oxford University Press}
}

@article{jiang2025embers,
  title={Embers of Active Galactic Nuclei: Tidal Disruption Events and Quasiperiodic Eruptions},
  author={Jiang, Ning and Pan, Zhen},
  journal={The Astrophysical Journal Letters},
  volume={983},
  number={1},
  pages={L18},
  year={2025},
  publisher={IOP Publishing}
}

@ARTICLE{vurm2025radiation,
       author = {{Vurm}, Indrek and {Linial}, Itai and {Metzger}, Brian D.},
        title = "{Radiation Transport Simulations of Quasiperiodic Eruptions from Star{\textendash}Disk Collisions}",
      journal = {\apj},
         year = 2025,
        month = apr,
       volume = {983},
       number = {1},
          eid = {40},
        pages = {40},
          doi = {10.3847/1538-4357/adb74d},
archivePrefix = {arXiv},
       eprint = {2410.05166},
 primaryClass = {astro-ph.HE},
       adsurl = {https://ui.adsabs.harvard.edu/abs/2025ApJ...983...40V}
}

@article{chakraborty2025discovery,
  title={Discovery of Quasiperiodic Eruptions in the Tidal Disruption Event and Extreme Coronal Line Emitter AT2022upj: Implications for the QPE/TDE Fraction and a Connection to ECLEs},
  author={Chakraborty, Joheen and Kara, Erin and Arcodia, Riccardo and Buchner, Johannes and Giustini, Margherita and Hern{\'a}ndez-Garc{\'\i}a, Lorena and Linial, Itai and Masterson, Megan and Miniutti, Giovanni and Mummery, Andrew and others},
  journal={The Astrophysical Journal Letters},
  volume={983},
  number={2},
  pages={L39},
  year={2025},
  publisher={IOP Publishing}
}

@article{chakraborty2024testing,
  title={Testing EMRI Models for Quasi-periodic Eruptions with 3.5 yr of Monitoring eRO-QPE1},
  author={Chakraborty, Joheen and Arcodia, Riccardo and Kara, Erin and Miniutti, Giovanni and Giustini, Margherita and Tetarenko, Alexandra J and Rhodes, Lauren and Franchini, Alessia and Bonetti, Matteo and Burdge, Kevin B and others},
  journal={The Astrophysical Journal},
  volume={965},
  number={1},
  pages={12},
  year={2024},
  publisher={IOP Publishing}
}

@article{arcodia2024more,
  title={The more the merrier: SRG/eROSITA discovers two further galaxies showing X-ray quasi-periodic eruptions},
  author={Arcodia, R and Liu, Z and Merloni, A and Malyali, A and Rau, A and Chakraborty, J and Goodwin, A and Buckley, D and Brink, J and Gromadzki, M and others},
  journal={Astronomy \& Astrophysics},
  volume={684},
  pages={A64},
  year={2024},
  publisher={EDP Sciences}
}

@article{arcodia2025srg,
  title={SRG/eROSITA No. 5: Discovery of quasi-periodic eruptions every\~{} 3.7 days from a galaxy at z> 0.1},
  author={Arcodia, R and Baldini, P and Merloni, A and Rau, A and Nandra, K and Chakraborty, J and Goodwin, AJ and Page, MJ and Buchner, J and Masterson, M and others},
  journal={arXiv preprint arXiv:2506.17138},
  year={2025}
}

@article{nicholl2024quasi,
  title={Quasi-periodic X-ray eruptions years after a nearby tidal disruption event},
  author={Nicholl, M and Pasham, DR and Mummery, A and Guolo, M and Gendreau, K and Dewangan, GC and Ferrara, EC and Remillard, R and Bonnerot, C and Chakraborty, J and others},
  journal={Nature},
  pages={1--5},
  year={2024},
  publisher={Nature Publishing Group UK London}
}

@article{arcodia2021x,
  title={X-ray quasi-periodic eruptions from two previously quiescent galaxies},
  author={Arcodia, R and Merloni, A and Nandra, K and Buchner, J and Salvato, M and Pasham, D and Remillard, R and Comparat, J and Lamer, G and Ponti, G and others},
  journal={Nature},
  volume={592},
  number={7856},
  pages={704--707},
  year={2021},
  publisher={Nature Publishing Group UK London}
}

@ARTICLE{ZhuStone2018,
       author = {{Zhu}, Zhaohuan and {Stone}, James M.},
        title = "{Global Evolution of an Accretion Disk with a Net Vertical Field: Coronal Accretion, Flux Transport, and Disk Winds}",
      journal = {\apj},
         year = 2018,
        month = apr,
       volume = {857},
       number = {1},
          eid = {34},
        pages = {34},
          doi = {10.3847/1538-4357/aaafc9},
archivePrefix = {arXiv},
       eprint = {1701.04627},
 primaryClass = {astro-ph.EP},
       adsurl = {https://ui.adsabs.harvard.edu/abs/2018ApJ...857...34Z}
}

@article{lu2023quasi,
  title={Quasi-periodic eruptions from mildly eccentric unstable mass transfer in galactic nuclei},
  author={Lu, Wenbin and Quataert, Eliot},
  journal={Monthly Notices of the Royal Astronomical Society},
  volume={524},
  number={4},
  pages={6247--6266},
  year={2023},
  publisher={Oxford University Press}
}

@article{miniutti2019nine,
  title={Nine-hour X-ray quasi-periodic eruptions from a low-mass black hole galactic nucleus},
  author={Miniutti, G and Saxton, RD and Giustini, M and Alexander, KD and Fender, RP and Heywood, I and Monageng, I and Coriat, M and Tzioumis, AK and Read, AM and others},
  journal={Nature},
  volume={573},
  number={7774},
  pages={381--384},
  year={2019},
  publisher={Nature Publishing Group UK London}
}

@article{xian2021x,
  title={X-ray quasi-periodic eruptions driven by star--disk collisions: application to GSN069 and probing the spin of massive black holes},
  author={Xian, Jingtao and Zhang, Fupeng and Dou, Liming and He, Jiasheng and Shu, Xinwen},
  journal={The Astrophysical Journal Letters},
  volume={921},
  number={2},
  pages={L32},
  year={2021},
  publisher={IOP Publishing}
}

@article{zhou2024probing,
  title={Probing orbits of stellar mass objects deep in galactic nuclei with quasi-periodic eruptions--III: Long term evolution},
  author={Zhou, Cong and Zeng, Yuhe and Pan, Zhen},
  journal={arXiv preprint arXiv:2411.18046},
  year={2024}
}

@article{xian2025secular,
  title={The Secular Periodic Evolution of X-ray Quasi-periodic Eruptions Driven by Star-disc Collisions},
  author={Xian, Jiajun and Zhang, Fupeng and Dou, Liming and Chen, Zhining},
  journal={arXiv preprint arXiv:2505.02596},
  year={2025}
}

@article{franchini2023quasi,
  title={Quasi-periodic eruptions from impacts between the secondary and a rigidly precessing accretion disc in an extreme mass-ratio inspiral system},
  author={Franchini, Alessia and Bonetti, Matteo and Lupi, Alessandro and Miniutti, Giovanni and Bortolas, Elisa and Giustini, Margherita and Dotti, Massimo and Sesana, Alberto and Arcodia, Riccardo and Ryu, Taeho},
  journal={Astronomy \& Astrophysics},
  volume={675},
  pages={A100},
  year={2023},
  publisher={EDP Sciences}
}

@article{raj2021disk,
  title={Disk tearing: implications for black hole accretion and AGN variability},
  author={Raj, Anagha and Nixon, CJ},
  journal={The Astrophysical Journal},
  volume={909},
  number={1},
  pages={82},
  year={2021},
  publisher={IOP Publishing}
}

@article{pan2022disk,
  title={A disk instability model for the quasi-periodic eruptions of GSN 069},
  author={Pan, Xin and Li, Shuang-Liang and Cao, Xinwu and Miniutti, Giovanni and Gu, Minfeng},
  journal={The Astrophysical Journal Letters},
  volume={928},
  number={2},
  pages={L18},
  year={2022},
  publisher={IOP Publishing}
}

@article{kaur2023magnetically,
  title={Magnetically dominated discs in tidal disruption events and quasi-periodic eruptions},
  author={Kaur, Karamveer and Stone, Nicholas C and Gilbaum, Shmuel},
  journal={Monthly Notices of the Royal Astronomical Society},
  volume={524},
  number={1},
  pages={1269--1290},
  year={2023},
  publisher={Oxford University Press}
}

@article{krolik2022quasiperiodic,
  title={Quasiperiodic erupters: a stellar mass-transfer model for the radiation},
  author={Krolik, Julian H and Linial, Itai},
  journal={The Astrophysical Journal},
  volume={941},
  number={1},
  pages={24},
  year={2022},
  publisher={IOP Publishing}
}

@article{zhao2022quasi,
  title={Quasi-periodic eruptions from the helium envelope of hydrogen-deficient stars stripped by supermassive black holes},
  author={Zhao, ZY and Wang, YY and Zou, YC and Wang, FY and Dai, ZG},
  journal={Astronomy \& Astrophysics},
  volume={661},
  pages={A55},
  year={2022},
  publisher={EDP Sciences}
}

@ARTICLE{Linial2024coupled,
       author = {{Linial}, Itai and {Metzger}, Brian D.},
        title = "{Coupled Disk-star Evolution in Galactic Nuclei and the Lifetimes of QPE Sources}",
      journal = {\apj},
         year = 2024,
        month = oct,
       volume = {973},
       number = {2},
          eid = {101},
        pages = {101},
          doi = {10.3847/1538-4357/ad639e},
archivePrefix = {arXiv},
       eprint = {2404.12421},
 primaryClass = {astro-ph.HE},
       adsurl = {https://ui.adsabs.harvard.edu/abs/2024ApJ...973..101L}
}

@article{linial2023unstable,
  title={Unstable mass transfer from a main-sequence star to a supermassive black hole and quasiperiodic eruptions},
  author={Linial, Itai and Sari, Re’em},
  journal={The Astrophysical Journal},
  volume={945},
  number={2},
  pages={86},
  year={2023},
  publisher={IOP Publishing}
}

@article{giustini2020x,
  title={X-ray quasi-periodic eruptions from the galactic nucleus of RX J1301. 9+ 2747},
  author={Giustini, Margherita and Miniutti, Giovanni and Saxton, Richard D},
  journal={Astronomy \& Astrophysics},
  volume={636},
  pages={L2},
  year={2020},
  publisher={EDP Sciences}
}

@article{chakraborty2021possible,
  title={Possible X-ray quasi-periodic eruptions in a tidal disruption event candidate},
  author={Chakraborty, Joheen and Kara, Erin and Masterson, Megan and Giustini, Margherita and Miniutti, Giovanni and Saxton, Richard},
  journal={The Astrophysical Journal Letters},
  volume={921},
  number={2},
  pages={L40},
  year={2021},
  publisher={IOP Publishing}
}

@ARTICLE{Quintin2023,
       author = {{Quintin}, E. and {Webb}, N.~A. and {Guillot}, S. and {Miniutti}, G. and {Kammoun}, E.~S. and {Giustini}, M. and {Arcodia}, R. and {Soucail}, G. and {Clerc}, N. and {Amato}, R. and {Markwardt}, C.~B.},
        title = "{Tormund's return: Hints of quasi-periodic eruption features from a recent optical tidal disruption event}",
      journal = {\aap},
         year = 2023,
        month = jul,
       volume = {675},
          eid = {A152},
        pages = {A152},
          doi = {10.1051/0004-6361/202346440},
archivePrefix = {arXiv},
       eprint = {2306.00438},
 primaryClass = {astro-ph.HE},
       adsurl = {https://ui.adsabs.harvard.edu/abs/2023A&A...675A.152Q}
}

@ARTICLE{Arcodia2022ero1,
       author = {{Arcodia}, R. and {Miniutti}, G. and {Ponti}, G. and {Buchner}, J. and {Giustini}, M. and {Merloni}, A. and {Nandra}, K. and {Vincentelli}, F. and {Kara}, E. and {Salvato}, M. and {Pasham}, D.},
        title = "{The complex time and energy evolution of quasi-periodic eruptions in eRO-QPE1}",
      journal = {\aap},
         year = 2022,
        month = jun,
       volume = {662},
          eid = {A49},
        pages = {A49},
          doi = {10.1051/0004-6361/202243259},
archivePrefix = {arXiv},
       eprint = {2203.11939},
 primaryClass = {astro-ph.HE},
       adsurl = {https://ui.adsabs.harvard.edu/abs/2022A&A...662A..49A}
}

@ARTICLE{HernandezGarcia2025,
       author = {{Hern{\'a}ndez-Garc{\'\i}a}, Lorena and {Chakraborty}, Joheen and {S{\'a}nchez-S{\'a}ez}, Paula and {Ricci}, Claudio and {Cuadra}, Jorge and {McKernan}, Barry and {Ford}, K.~E. Saavik and {Ar{\'e}valo}, Patricia and {Rau}, Arne and {Arcodia}, Riccardo and {Kara}, Erin and {Liu}, Zhu and {Merloni}, Andrea and {Bruni}, Gabriele and {Goodwin}, Adelle and {Arzoumanian}, Zaven and {Assef}, Roberto J. and {Baldini}, Pietro and {Bayo}, Amelia and {Bauer}, Franz E. and {Bernal}, Santiago and {Brightman}, Murray and {Calistro Rivera}, Gabriela and {Gendreau}, Keith and {Homan}, David and {Krumpe}, Mirko and {Lira}, Paulina and {Mart{\'\i}nez-Aldama}, Mary Loli and {Salvato}, Mara and {Sotomayor}, Bel{\'e}n},
        title = "{Discovery of extreme quasi-periodic eruptions in a newly accreting massive black hole}",
      journal = {Nature Astronomy},
         year = 2025,
        month = apr,
          doi = {10.1038/s41550-025-02523-9},
archivePrefix = {arXiv},
       eprint = {2504.07169},
 primaryClass = {astro-ph.HE},
       adsurl = {https://ui.adsabs.harvard.edu/abs/2025NatAs.tmp...90H}
}

@ARTICLE{Wevers2022,
       author = {{Wevers}, T. and {Pasham}, D.~R. and {Jalan}, P. and {Rakshit}, S. and {Arcodia}, R.},
        title = "{Host galaxy properties of quasi-periodically erupting X-ray sources}",
      journal = {\aap},
         year = 2022,
        month = mar,
       volume = {659},
          eid = {L2},
        pages = {L2},
          doi = {10.1051/0004-6361/202243143},
archivePrefix = {arXiv},
       eprint = {2201.11751},
 primaryClass = {astro-ph.HE},
       adsurl = {https://ui.adsabs.harvard.edu/abs/2022A&A...659L...2W}
}

@ARTICLE{Wang2024,
       author = {{Wang}, Yihan and {Zhu}, Zhaohuan and {Lin}, Douglas N.~C.},
        title = "{Stellar/BH population in AGN discs: direct binary formation from capture objects in nuclei clusters}",
      journal = {\mnras},
         year = 2024,
        month = mar,
       volume = {528},
       number = {3},
        pages = {4958-4975},
          doi = {10.1093/mnras/stae321},
archivePrefix = {arXiv},
       eprint = {2308.09129},
 primaryClass = {astro-ph.GA},
       adsurl = {https://ui.adsabs.harvard.edu/abs/2024MNRAS.528.4958W}
}

@article{zalamea2010white,
  title={White dwarfs stripped by massive black holes: sources of coincident gravitational and electromagnetic radiation},
  author={Zalamea, Ivan and Menou, Kristen and Beloborodov, Andrei M},
  journal={Monthly Notices of the Royal Astronomical Society: Letters},
  volume={409},
  number={1},
  pages={L25--L29},
  year={2010},
  publisher={Blackwell Publishing Ltd Oxford, UK}
}

@article{king2020gsn,
  title={GSN 069--a tidal disruption near miss},
  author={King, Andrew},
  journal={Monthly Notices of the Royal Astronomical Society: Letters},
  volume={493},
  number={1},
  pages={L120--L123},
  year={2020},
  publisher={Oxford University Press}
}

@article{king2022quasi,
  title={Quasi-periodic eruptions from galaxy nuclei},
  author={King, Andrew},
  journal={Monthly Notices of the Royal Astronomical Society},
  volume={515},
  number={3},
  pages={4344--4349},
  year={2022},
  publisher={Oxford University Press}
}

@article{rom2024dynamics,
  title={Dynamics Around Supermassive Black Holes: Extreme-mass-ratio Inspirals as Gravitational-wave Sources},
  author={Rom, Barak and Linial, Itai and Kaur, Karamveer and Sari, Re’em},
  journal={The Astrophysical Journal},
  volume={977},
  number={1},
  pages={7},
  year={2024},
  publisher={IOP Publishing}
}

@ARTICLE{Guo2026,
       author = {{Guo}, Hengxiao and {Yan}, Zhen and {Li}, Ya-Ping and {Chakraborty}, Joheen and {S{\'a}nchez-S{\'a}ez}, Paula and {Hern{\'a}ndez-Garc{\'\i}a}, Lorena and {Zhang}, Wenda and {Sun}, Jingbo and {Li}, Shuang-Liang and {Deng}, Hongping and {Zuo}, Wenwen and {Tagawa}, Hiromichi and {Pan}, Xin and {Zhang}, Minghao and {Ar{\'e}valo}, Patricia and {Lira}, Paulina and {Jin}, Chichuan and {Gu}, Minfeng},
        title = "{Evidence for a Delayed Ultraviolet Counterpart to X-Ray Quasiperiodic Eruptions in Ansky}",
      journal = {\apjl},
         year = 2026,
        month = apr,
       volume = {1000},
       number = {2},
          eid = {L57},
        pages = {L57},
          doi = {10.3847/2041-8213/ae524b},
archivePrefix = {arXiv},
       eprint = {2603.02517},
 primaryClass = {astro-ph.HE},
       adsurl = {https://ui.adsabs.harvard.edu/abs/2026ApJ..1000L..57G}
}

@article{sari2019tidal,
  title={Tidal disruption events, main-sequence extreme-mass ratio inspirals, and binary star disruptions in galactic nuclei},
  author={Sari, Re’em and Fragione, Giacomo},
  journal={The Astrophysical Journal},
  volume={885},
  number={1},
  pages={24},
  year={2019},
  publisher={IOP Publishing}
}

@article{mummery2025collisions,
  title={Collisions with tidal disruption event disks: implications for quasi-periodic X-ray eruptions},
  author={Mummery, Andrew},
  journal={arXiv preprint arXiv:2504.21456},
  year={2025}
}

@article{dodd2025perturbing,
  title={Perturbing AGN Accretion Disks with Stars and Moderately Massive Black Holes: Implications for Changing-Look AGN and Quasi-Periodic Eruptions},
  author={Dodd, Sierra A and Huang, Xiaoshan and Davis, Shane W and Ramirez-Ruiz, Enrico},
  journal={arXiv preprint arXiv:2506.19900},
  year={2025}
}

@article{goodwin2025radio,
  title={The radio properties of quasi-periodic X-ray eruption sources},
  author={Goodwin, AJ and Arcodia, R and Miniutti, G and Miller-Jones, JC and van Velzen, S},
  journal={arXiv preprint arXiv:2506.14417},
  year={2025}
}

@article{wevers2025time,
  title={Time-resolved Hubble Space Telescope UV Observations of an X-Ray Quasiperiodic Eruption Source},
  author={Wevers, Thomas and Guolo, Muryel and Lockwood, Sean and Mummery, Andrew and Pasham, Dheeraj R and Arcodia, Riccardo},
  journal={The Astrophysical Journal Letters},
  volume={980},
  number={1},
  pages={L1},
  year={2025},
  publisher={IOP Publishing}
}

@article{nicholl2020outflow,
  title={An outflow powers the optical rise of the nearby, fast-evolving tidal disruption event AT2019qiz},
  author={Nicholl, M and Wevers, T and Oates, SR and Alexander, KD and Leloudas, G and Onori, F and Jerkstrand, Anders and Gomez, S and Campana, S and Arcavi, I and others},
  journal={Monthly Notices of the Royal Astronomical Society},
  volume={499},
  number={1},
  pages={482--504},
  year={2020},
  publisher={Oxford University Press}
}

@article{newsome2024mapping,
  title={Mapping the Inner 0.1 pc of a Supermassive Black Hole Environment with the Tidal Disruption Event and Extreme Coronal-line Emitter AT 2022upj},
  author={Newsome, Megan and Arcavi, Iair and Howell, D Andrew and McCully, Curtis and Terreran, Giacomo and Hosseinzadeh, Griffin and Bostroem, K Azalee and Dgany, Yael and Farah, Joseph and Faris, Sara and others},
  journal={The Astrophysical Journal},
  volume={977},
  number={2},
  pages={258},
  year={2024},
  publisher={IOP Publishing}
}

@article{zhu2025ultraviolet,
  title={Ultraviolet Spectral Evidence for Ansky as a Slowly Evolving Featureless Tidal Disruption Event with Quasi-periodic Eruptions},
  author={Zhu, Jiazheng and Jiang, Ning and Wang, Yibo and Wang, Tinggui and Sun, Luming and Zhong, Shiyan and Yao, Yuhan and Chornock, Ryan and Dai, Lixin and Lyu, Jianwei and others},
  journal={arXiv preprint arXiv:2510.22211},
  year={2025}
}

@article{hernandez2025nicer,
  title={NICER observations reveal doubled timescales in Ansky's quasi-periodic eruptions (QPEs)},
  author={Hern{\'a}ndez-Garc{\'\i}a, L and S{\'a}nchez-S{\'a}ez, P and Chakraborty, J and Cuadra, J and Miniutti, G and Arcodia, R and Ar{\'e}valo, P and Giustini, M and Kara, E and Ricci, C and others},
  journal={arXiv preprint arXiv:2509.16304},
  year={2025}
}

@article{linial2025qpes,
  title={QPEs from EMRI Debris Streams Impacting Accretion Disks in Galactic Nuclei},
  author={Linial, Itai and Metzger, Brian D and Quataert, Eliot},
  journal={arXiv preprint arXiv:2506.10096},
  year={2025}
}

@article{suzuguchi2025quasi,
  title={Quasi-Periodic Eruptions as a Probe of Accretion Disk in Tidal Disruption Events},
  author={Suzuguchi, Tomoya and Matsumoto, Tatsuya},
  journal={arXiv preprint arXiv:2509.01663},
  year={2025}
}

@article{huang2025multi,
  title={Multi-band Emission from Star-Disk Collision and Implications for Quasi-Periodic Eruptions},
  author={Huang, Xiaoshan and Linial, Itai and Jiang, Yan-Fei},
  journal={arXiv preprint arXiv:2506.11231},
  year={2025}
}

@article{chakraborty2025prospects,
  title={Prospects for EMRI/MBH Parameter Estimation Using Quasiperiodic Eruption Timings: Short-timescale Analysis},
  author={Chakraborty, Joheen and Drummond, Lisa V and Bonetti, Matteo and Franchini, Alessia and Kejriwal, Shubham and Miniutti, Giovanni and Arcodia, Riccardo and Hughes, Scott A and Duque, Francisco and Kara, Erin and others},
  journal={The Astrophysical Journal},
  volume={992},
  number={1},
  pages={120},
  year={2025},
  publisher={IOP Publishing}
}

@article{naoz2025triples,
  title={Triples as Links between Binary Black Hole Mergers, Their Electromagnetic Counterparts, and Galactic Black Holes},
  author={Naoz, Smadar and Haiman, Zolt{\'a}n and Quataert, Eliot and Holzknecht, Liz},
  journal={The Astrophysical Journal Letters},
  volume={992},
  number={1},
  pages={L12},
  year={2025},
  publisher={IOP Publishing}
}

@article{short2023delayed,
  title={Delayed appearance and evolution of coronal lines in the TDE AT2019qiz},
  author={Short, P and Lawrence, A and Nicholl, M and Ward, M and Reynolds, TM and Mattila, S and Yin, C and Arcavi, I and Carnall, A and Charalampopoulos, P and others},
  journal={Monthly Notices of the Royal Astronomical Society},
  volume={525},
  number={1},
  pages={1568--1587},
  year={2023},
  publisher={Oxford University Press}
}

@ARTICLE{Yao2026,
       author = {{Yao}, Philippe Z. and {Quataert}, Eliot},
        title = "{Mass Transfer in Tidally Heated Stars Orbiting Massive Black Holes and Implications for Repeating Nuclear Transients}",
      journal = {The Open Journal of Astrophysics},
         year = 2026,
        month = jun,
       volume = {9},
        pages = {64077},
          doi = {10.33232/001c.164077},
archivePrefix = {arXiv},
       eprint = {2505.10611},
 primaryClass = {astro-ph.HE},
       adsurl = {https://ui.adsabs.harvard.edu/abs/2026OJAp....964077Y}
}

@ARTICLE{Linial2025,
       author = {{Linial}, Itai and {Metzger}, Brian D. and {Quataert}, Eliot},
        title = "{QPEs from EMRI Debris Streams Impacting Accretion Disks in Galactic Nuclei}",
      journal = {\apj},
         year = 2025,
        month = oct,
       volume = {991},
       number = {2},
          eid = {147},
        pages = {147},
          doi = {10.3847/1538-4357/adfa0e},
archivePrefix = {arXiv},
       eprint = {2506.10096},
 primaryClass = {astro-ph.HE},
       adsurl = {https://ui.adsabs.harvard.edu/abs/2025ApJ...991..147L}
}

@ARTICLE{Jankovic2026,
       author = {{Jankovi{\v{c}}}, Taj and {Karpov}, Sergey and {Zaja{\v{c}}ek}, Michal and {Karas}, Vladim{\'\i}r and {{\'S}niegowska}, Marzena},
        title = "{Radiation-hydrodynamics of star-disc collisions: From system parameters to outflows and lightcurves}",
      journal = {arXiv e-prints},
         year = 2026,
        month = jul,
          eid = {arXiv:2607.05508},
        pages = {arXiv:2607.05508},
          doi = {10.48550/arXiv.2607.05508},
archivePrefix = {arXiv},
       eprint = {2607.05508},
 primaryClass = {astro-ph.HE},
       adsurl = {https://ui.adsabs.harvard.edu/abs/2026arXiv260705508J}
}

@ARTICLE{Nakar2010,
       author = {{Nakar}, Ehud and {Sari}, Re'em},
        title = "{Early Supernovae Light Curves Following the Shock Breakout}",
      journal = {\apj},
         year = 2010,
        month = dec,
       volume = {725},
       number = {1},
        pages = {904-921},
          doi = {10.1088/0004-637X/725/1/904},
archivePrefix = {arXiv},
       eprint = {1004.2496},
 primaryClass = {astro-ph.HE},
       adsurl = {https://ui.adsabs.harvard.edu/abs/2010ApJ...725..904N}
}

@article{jankovivc2026radiation,
  title={Radiation-hydrodynamics of star-disc collisions for quasi-periodic eruptions},
  author={Jankovi{\v{c}}, Taj and Bonnerot, Cl{\'e}ment and Karpov, Sergey and Jurca, Aleksej},
  journal={arXiv preprint arXiv:2602.02656},
  year={2026}
}

@article{liu2026quasi,
  title={Quasi-periodic Eruptions from Stellar-mass Black Holes Impacting Accretion Disks in Galactic Nuclei},
  author={Liu, Kun and Liu, Shang-Fei and Pan, Zhen and Deng, Hongping and Shen, Rongfeng and Yu, Cong},
  journal={arXiv preprint arXiv:2603.00226},
  year={2026}
}
\bibliographystyle{aasjournalv7}



\end{CJK*}
\end{document}